\documentclass[11pt,authoryear]{elsarticle}
\usepackage[margin=2.3cm, includehead]{geometry}

\usepackage{currfile}
\usepackage{mathrsfs}
\usepackage[mathscr]{eucal}
\DeclareMathAlphabet{\mathpzc}{OT1}{pzc}{m}{it}
\usepackage{natbib}
\usepackage{makeidx}
\usepackage{multirow}
\usepackage{multicol}
\usepackage{booktabs}
\usepackage{hhline}
\usepackage[dvipsnames,svgnames,table]{xcolor}
\usepackage{graphicx}
\usepackage{epstopdf}
\usepackage{ulem}
\usepackage{amsmath}
\usepackage{amsfonts}
\usepackage{url}
\usepackage{amssymb}
\usepackage{graphics}
\usepackage{epsfig}
\usepackage{verbatim}
\usepackage{amsthm}
\usepackage{geometry}
\usepackage{framed}    
\usepackage{scalefnt}
\usepackage{alltt}

\def\i.i.d.{\buildrel {\rm i.i.d.} \over \sim}

\def\cw#1 { \overset{\mathbb{P}}{\underset{#1}{\longrightarrow}} }
\def\Real{\mathbb{R}}
\def\Natu0{\mathbb{N}_0}

\def\E#1{{\mathrm E}\left[#1\right]}

\def\Var#1{{\mathrm Var}\left(#1\right)}

\def \rcov#1#2 {{\rm cov}_{#1}\left( #2\right)}

\DeclareMathOperator*{\argmin}{arg\,min}

\newtheorem{example}{Example}

\newtheorem{lemma}{Lemma}
\newtheorem{theorem}{Theorem}
\newtheorem{definition}{Definition}
\newtheorem{corollary}{Corollary}
\newtheorem{remark}{Remark}

\newtheorem{model}{Model}

\def\cov#1{{\rm  Cov}\left[#1\right]}

\begin{document}
	\begin{titlepage}
		
		\thispagestyle{empty}
\title{On the ordering of credibility factors
}

%
%
%
%
%

\author{Jae Youn Ahn\corref{bbb}\fnref{thirdfoot}}
\author{Himchan Jeong\corref{bbb}\fnref{firstfoot}}
\author{Yang Lu\corref{bbb}\fnref{secondfoot}}

\cortext[bbb]{Corresponding Authors}
\fntext[thirdfoot]{Department of Statistics, Ewha Womans University, Seoul, Republic of Korea. Email: \url{jaeyahn@ewha.ac.kr}}
\fntext[firstfoot]{Department of Statistics and Actuarial Science, Simon Fraser University, BC, Canada. Email: \url{himchan_jeong@sfu.ca}}
\fntext[secondfoot]{Department of Mathematics and Statistics, Concordia University, Montreal, QC Canada. Email: \url{yang.lu@concordia.ca}}


\begin{abstract}
Traditional credibility analysis of risks in insurance is based on the random effects model, where the heterogeneity across the policyholders is assumed to be time-invariant. One popular extension is the dynamic random effects (or state-space) model. However, while the latter allows for time-varying heterogeneity, its application to the credibility analysis should be conducted with care due to the possibility of negative credibilities per period [see \citet{pinquet2020poisson}].
Another important but under-explored topic is the ordering of the credibility factors in a monotonous manner---recent claims ought to have larger weights than the old ones. This paper shows that the ordering of the covariance structure of the random effects in the dynamic random effects model does not necessarily imply that of the credibility factors. Subsequently, we show that the state-space model, with AR(1)-type autocorrelation function, guarantees the ordering of the credibility factors. Simulation experiments and a case study with a real dataset are conducted to show the relevance in insurance applications.

\end{abstract}


\end{titlepage}


\maketitle




\vfill

\pagebreak

\section{Introduction}

Insurers use credibility theory to adjust a policyholder's premium based on past claim history. Typically, the a priori risk premium is first determined by the policyholder's observable risk characteristics. Then the a posteriori premium is expressed as weighted sums of the past claims, and these weights are called credibility factors.
The baseline model that underlies many of the early contributions in credibility theory is the static random effects model.
In this case, a posteriori premium is expressed as a weighted sum of two terms: the sample mean and a priori risk premium. Consequently, a posteriori premium imposes the same credibility factors to each claim regardless of its seniority. Motivated by this downside of the time-invariant credibility factors, the literature has considered the age of claims by imposing geometric decreasing credibility factors \citep{gerber1973credibility, gerber1975credibility, sundt1988credibility}.\footnote{These papers, however, do not allow for individual-specific and potentially time-varying risk exposure and are thus unsuitable for many real applications.}




An alternative way to account for the seniority of the claims is through the introduction of the dynamic random effects rather than static random effects [\citet{bolance2003time}]. Subsequently, the seniority of the claims is indirectly considered by allowing the dependence between random effects to decrease as time lag increases.
These models fall within the scope of state-space models (SSM) and are computationally intensive to deal with.  In particular, they do not allow for closed-form conditional expectation in general, which may make them computationally inconvenient for a posteriori ratemaking [e.g., \citet{brouhns2003bonus} and \citet{li2020dynamic}].  Furthermore, computing the conditional expectation requires the full specification of the dynamics of the unobserved random effect process,  which may give rise to more mis-specification risk.
Consequently, the forecasting in the dynamic random effect model is often approximated by the credibility premium. However, not all dynamic random effects models lead to sensible credibility premium in the insurance perspective. 
Recently, \citet{pinquet2020poisson} shows that the credibility coefficients associated with these models are not necessarily ensured to be non-negative. More precisely, \citet{pinquet2020poisson} shows that in Poisson dynamic random effects models where the random effect has an AR($p$)-type auto-covariance function, the non-negativity condition is not satisfied without parametric constraints on the auto-covariance function.\footnote{More recently, in a follow-up work, \cite{pinquet2020positivity} shows that an ARFIMA(0,d,0) type auto-correlation of the dynamic random effects is compatible with the positivity of the credibility coefficients. However, the latter paper does not provide examples of \textit{positive} processes with ARFIMA(0,d,0) representation.} This result has important implications as it demonstrates the limitations of the credibility approach when used semi-parametrically, that is when \textit{actuarial meaningfulness} is not considered.


This study furthers this investigation of dynamic random effects models, in a similar spirit as \cite{pinquet2020poisson}. We focus on another important but thus far overlooked property of whether the credibility factors are decreasing in seniority. While this property is a motivation of the dynamic random effects in actuarial perspective [e.g., \cite{pinquet2001allowance, bolance2003time, lu2018dynamic}], until now, it has not been \textit{formally} checked in the literature.
We start our analysis with a simple question: does a decreasing auto-correlation function (ACF) of the dynamic random effect suffice for the resulting credibility coefficients to be also decreasing in seniority? We show that this is generically \textit{not} the case.
This leads us to restrict our attention to dynamic random effect processes with AR(1) autocorrelation such as those introduced by \citep{grunwald2000theory}. Under such an auto-correlation restriction, we show that the corresponding credibility factor is an increasing function of time as well as being non-negative.


The paper is organized as follows. Section 2 provides a brief introduction to the dynamic random effects models and credibility theory. In particular, we demonstrate that the ordering of the covariance does not always guarantee that of the credibility factor. Section \ref{sec.3} provides the specific formulation of the dynamic random effects model, which is guaranteed to have the ordered credibility factors. In Section 4, we perform a numerical study under various parameter settings. In Section 5, an empirical analysis is conducted on a real-life dataset.

\section{The general setting and motivating examples}
This section provides a quick reminder of the general setting, the definition of the credibility coefficients, and two motivating examples for which the credibility coefficients are not monotonously ordered.
\subsection{Notation and definition}
 Let us assume that for each individual $i$, we observe claim-related variables $Y_{i,t}$ for policy year $t=1,...,T_i$. We denote the history of claims up to time $T_i$ by
$$
\mathcal{F}_{i,T_i} := \left\{ Y_{i,t}\, \big| \,  t=1,\ldots,T_i,  \right \},
$$
and assume that the sequence $(Y_{i,t}: t=1, 2, \cdots)$ is independent across different individuals $i$.
In posterior ratemaking, it is interesting to predict $Y_{i,T_i+1}$ for each policyholder $i$ and upcoming year $T_i+1$, which is given as
\begin{equation}\label{exact.1}
\E{Y_{i, T_i+1}| \mathcal{F}_{i,T_i}}.
\end{equation}
For the expository purpose, from now onward, we remove subscript $i$.

We say that a sequence of stationary random variables $(Y_t:t=1,2, \cdots)$ has {\it positive covariance ordering} if the autocovariance function (ACF) is decreasing in lag:
\begin{equation}\label{eq.2}
\cov{Y_t, Y_{t+p}}>\cov{Y_t, Y_{t+q}}, \quad\hbox{for all}\quad p,q,t=1, 2, \cdots
\end{equation}
satisfying $p<q$.

For the distributional assumption of the observed claims, we consider the \textit{reproductive exponential dispersion family} (EDF) by \citet{Nelder1989}.  They show a real-valued random variable ${\displaystyle Y}$ as belonging to the reproductive exponential dispersion model with mean parameter $u$ and dispersion parameter $\psi$ denoted by ${\rm ED}\left(u, \psi \right)$ and has a density function at $Y=y$ of the following form
\[
 h(y; u, \psi)\exp\left(\frac{\theta y - A(\theta)}{\psi} \right).
\]
Here, $h$ is a suitable normalizing term, and $A$ is twice a continuously differentiable function satisfying
\begin{equation*}
\theta = \left(A^{\prime}\right)^{-1}(u).
\end{equation*}
If $Y\sim {\rm ED}\left(u, \psi \right)$, we have
\[
\E{Y}=u\quad\hbox{and}\quad\Var{Y}=V(u)\psi
\]
where $V$, called the {\it unit variance function}, is defined by
\[
V(u)=A^{\prime\prime}\left( \left( A^{\prime}\right)^{-1}(u)\right).
\]
This study utilizes the following examples of distributions in EDF:
\begin{itemize}
  \item ${\rm Pois}(\lambda)$: the Poisson distribution\footnote{For Poisson distribution, the corresponding dispersion parameter is $\psi=1$, and the corresponding unit variance function is given by $V(\mu)=\mu.$} with parameter $\lambda$;
  \item ${\rm Gamma}(\lambda, \psi)$: the gamma distribution\footnote{Here, $\psi$ is dispersion parameter, and the corresponding unit variance function is given by $V(\mu)=\mu^2$.} with mean $\lambda$ and variance $\lambda^2\psi$.
\end{itemize}

We use ${\rm Beta}(a, b)$ to denote the beta distribution with parameters $a, b$.

Finally, we also define three $T\times T$ matrices. We denote by $\boldsymbol{I}_T$ the identity matrix, $\boldsymbol{E}_T$ the matrix of 1, and $\boldsymbol{\Sigma}_{T, \rho}$ the Toeplitz matrix whose entries are given by
\[
\left[\boldsymbol{\Sigma}_{T, \rho}\right]_{i,j}=\rho^{|i-j|}, \quad i, j =1, \cdots, T.
\]
where $\rho \in (-1, 1)$. We recall the following well-known result concerning its inverse matrix:
\begin{equation}\label{eq.15}
   (1-\rho^2) \left(\boldsymbol{\Sigma}_{T, \rho}\right)^{-1} =
    \begin{cases}
      1, & (i,j)=(1,1), \quad\hbox{or}\quad (i,j)=(T,T);\\
      1+\rho^2, & (i,j)\neq(1,1), (i,j)\neq(T,T),\quad\hbox{and}\quad i=j;\\
      -\rho, & j=i+1\quad\hbox{or}\quad j=i-1;\\
      0, & \hbox{otherwise}.
    \end{cases}
\end{equation}


\subsection{Brief reminder of the credibility theory}

While the conditional expectation in \eqref{exact.1} is the best forecast in terms of the mean squared error (MSE), it can be problematic to communicate with the policyholders if it does not allow the closed form solution \citep{goulet1998principles}.
Alternatively, the credibility premium provides the efficient yet intuitive affine functional form of the premium, which facilitates communication with the policyholders.
Formally, for a sequence of random variables $\left( Y_t; t=1, 2, \cdots\right)$ satisfying $\E{Y_t}=\lambda_t$,
 the credibility premium for time $T+1$ is defined by
\begin{equation}\label{eq.10}
{\rm Prem}\left(Y_1, \cdots, Y_T\right):=\widehat{\alpha}_0\lambda_{T+1}+\widehat{\alpha}_1Y_1+\cdots+\widehat{\alpha}_TY_T
\end{equation}
where
\begin{equation}\label{eq.11}
\left(\widehat{\alpha}_0,\cdots, \widehat{\alpha}_T  \right):=\argmin\limits_{(\alpha_0, \cdots, \alpha_T)\in\Real^{T+1}}\E{\left(Y_{T+1} -
\left(\alpha_0\lambda_{T+1}+\alpha_1Y_1+\cdots+\alpha_TY_T\right)
\right)^2}.
\end{equation}
Here, $\widehat{\alpha}_t$ for $t=1, \cdots, T$ is referred to as the credibility factor, which can be interpreted as the contribution of $t$-th year observation to the credibility premium. This credibility premium, which is the best linear unbiased estimator (BLUE) of
$\E{Y_{T+1}|\mathcal{F}_T}$, can be computed through [see \citet{buhlmann2006course}]:
\begin{equation}\label{eq.22}
\left(\widehat{\alpha}_1,\cdots, \widehat{\alpha}_T  \right)'
=\boldsymbol{\Sigma}_T^{-1}\left(
\begin{array}{c}
  \cov{Y_1, Y_{T+1}} \\
  \cov{Y_2, Y_{T+1}} \\
  \vdots \\
  \cov{Y_T, Y_{T+1}}
\end{array}
\right)
\end{equation}
and
\begin{equation}\label{eq.33}
\widehat{\alpha}_0=\frac{\E{Y_{T+1}}-\sum\limits_{t=1}^{T}\widehat{\alpha}_t\E{Y_{t}}}{\lambda_{T+1}}
\end{equation}
where $\boldsymbol{\Sigma}_T$ is a  $T\times T$ covariance matrix  defined by
\[
\left[ \boldsymbol{\Sigma}_T \right]_{i,j}=
\cov{Y_i, Y_j}
\]

We say that the credibility premium is {\it regular} if
\[
\widehat{\alpha}_t>0, \quad \hbox{for}\quad t=1, \ldots.
\]
We note that negative credibility factors are highly undesirable for regulatory reasons, as they might give the wrong incentives of making claims to get lower premiums or even give rise to negative premiums, which is the synonym of arbitrage opportunities [e.g., \citet{pinquet2020poisson}, \citet{li2020dynamic}].
For the condition of the regular credibility premium, we refer to the study of \citet{pinquet2020poisson, pinquet2020positivity}.

The following example shows the property of credibility factors under the static random effects model.
\begin{example}[Static random effects model]

For a distribution function $G$, consider a random effect $R\sim G$ so that $Y_j$'s are conditionally independent for given $R$. Subsequently, after integrating the random effect, we obtain the joint  distribution of $(Y_1, \cdots, Y_T, Y_{T+1})$. However, random effects models may not be ideal for the posterior ratemaking in insurance products mainly due the symmetric dependence structure. Consequently, old claims are equally treated as new claims in the determination of future premiums, which is clearly counterintuitive.
More precisely, as the covariance matrix in \eqref{eq.22} has
the equi-covariance structure,\footnote{We assume that marginal distributions are the same.} the following identity holds:
\begin{equation}\label{eq.3}
\widehat{\alpha}_1=\cdots=\widehat{\alpha}_T.
\end{equation}
Clearly, the credibility premium having the credibility factors of the form in \eqref{eq.3} is counterintuitive when applied to insurance ratemaking,
in the sense that the weight of old claims, for example, $\widehat{\alpha}_1$, is the same as the weight of new claims, for example, $\widehat{\alpha}_T$.
\end{example}

More realistically, one may depart from the equi-covariance assumption and consider
a series of random variables $(Y_t:t=1,2, \cdots)$  specified as follows, which has the positive covariance ordering as in \eqref{eq.2}.

\begin{example}[Dynamic random effects model]

Let us consider a dynamic random effects (or state-space) model comprising an unobserved process $(R_t:t=0,1, \cdots)$ (called state variable) and an observed process $(Y_t:t=1,2, \cdots)$. We also assume that for $t$ varying, $Y_t$'s are independent conditionally on $(R_t:t=1,2, \cdots)$, and their distribution depends on $R_t$ only. Consequently, the joint density function of $\left(Y_{1:T}, R_{0:T}\right)$ 
 can be expressed as
\[
f\left(y_{1:T}, r_{0:T}\right)=f\left(r_0\right)\prod\limits_{t=1}^{T}f(y_t|r_t)f\left(r_t|r_{t-1}\right).
\]

In other words, such a model has the causal chain:
\begin{equation}
\label{chainstructure}
\begin{tabular}{ccccccccccccccccccccc}
$\cdots$     & $\longrightarrow$  & $R_{t-1}$   & $\longrightarrow$   & $R_{t}$    &  $\longrightarrow$  & $R_{t+1}$  & $\longrightarrow$  & $\cdots$ \\
$\downarrow$ &                 & $\downarrow$  &               &$\downarrow$   &                  & $\downarrow$  &             & $\downarrow$   \\
  $\cdots$ &                   & $Y_{t-1}$ &                    &$Y_{t}$      &                 & $Y_{t+1}$  &                  & $\cdots$
\end{tabular}
\end{equation}

Consequently, the specification of the random effects model $\left(Y_{1:T}, R_{0:T}\right)$ boils down to the specification of
\begin{enumerate}
  \item[i.] The dynamics of the sequence $(R_t:t=0,1, \ldots)$.
  \item[ii.] The conditional distribution of $Y_t$ for given $R_t$ for each $t$.
\end{enumerate}
\end{example}




Assuming that the sequence $(r_t:t=1,2, \cdots)$ has the positive covariance ordering,
it is natural to ask whether more recent observations have more influence than the old observations in the determination of the premium---whether we have
\begin{equation}\label{eq.16}
\widehat{\alpha}_1\le \cdots\le\widehat{\alpha}_T.
\end{equation}
under the framework of credibility theory. It will be shown in the next subsection that the
positive covariance ordering of $(Y_t)$ or $(R_t)$
is \textit{not} sufficient to guarantee \eqref{eq.16}.

\subsection{Some motivating state-space counter-examples} \label{sec.2.3}


This subsection provides two motivating examples of the stationary dynamic random effects model $(Y_t)$, which has a positive covariance ordered random effects $(R_t)$, without the credibility coefficients being monotonically decreasing in seniority.

\begin{example}[Semi-parametric model]

Let us first consider the first dynamic random effects model pioneered by \cite{bolance2003time}. This model is semi-parametric in the sense that the auto-covariance of the dynamic random effect process is not constrained. We set $\sigma^2=1$, and take the estimate of the auto-correlation function from their paper (see their Table 1):
$$
\rho_1=.733, \
\rho_2=.524, \
\rho_3=.504, \
\rho_4=.483, \
\rho_5=.401.
$$
Then we compute the credibility coefficients with time invariant $\lambda_t=1$, to obtain
the credibility coefficients which are equal to
$$
\begin{pmatrix}
\alpha_{3,T} \\
 \alpha_{2,T}  \\
   \alpha_{1,T}
\end{pmatrix}
=\begin{pmatrix}
 0.29 \\
 0.10 \\
  0.14
	\end{pmatrix}
$$
if $T=3$, or:
$$
\begin{pmatrix}
 \alpha_{4,T}\\
\alpha_{3,T} \\
 \alpha_{2,T}  \\
   \alpha_{1,T}
\end{pmatrix}
=\begin{pmatrix}
 0.28 \\
 0.09 \\
0.11 \\
 0.11
\end{pmatrix}
$$
if $T=4$, or:
$$
\begin{pmatrix}
  \alpha_{5,T}\\
 \alpha_{4,T}\\
\alpha_{3,T} \\
 \alpha_{2,T}  \\
   \alpha_{1,T}
\end{pmatrix}
=\begin{pmatrix}
 0.27 \\
 0.09 \\
0.10\\
0.09 \\
0.05
\end{pmatrix}
$$
if $T=5$. In particular, in none of the three cases, the credibility coefficients are correctly ordered.

\end{example}

\begin{example}[A parametric model]

The previous example shows the importance of putting constraints on the autocorrelation function of the dynamic random effect process rather than leaving it unconstrained. Let us now inspect a parametric model with a simple autocorrelation function.

Consider a state-space model that comprises a state process $(\boldsymbol{R}_t:t=0,1, \ldots)$ and observable time series $(Y_t:t=1,2, \ldots)$ of the following form:
  \begin{enumerate}
    \item[i.] The conditional distribution of $Y_t$ for given $\boldsymbol{R}_t:=\left( R_{t,1}, R_{t,2}\right)$ is
    \[
    Y_t|\boldsymbol{R}_t \sim {\rm ED}\left( \lambda_t \left({R_{t,1}+R_{t,2}}\right), \psi\right).
    \]
    \item[ii.] $(R_{t,1}:t=0,1, \ldots)$ and $(R_{t,2}:t=0,1, \ldots)$ are independent processes such that $(R_{t,1}:t=0,1, \ldots)$ has AR(1)-type autocorrelation:
  \begin{equation}\label{ex.eq.1}
  \E{R_{t,1}}=1, \quad \Var{R_{t,1}}=\sigma_1^2, \quad\hbox{and}\quad  \cov{R_{t_1,1}, R_{t_2,1}}=\sigma_1^2\rho^{|t_2-t_1|}, \quad t, t_1,t_2\in\mathbb{N}_0,
  \end{equation}
  and $R_{t,2}$ is time-invariant:
  \begin{equation}\label{ex.eq.2}
  \E{R_{t,2}}=1, \quad \Var{R_{t,2}}=\sigma_2^2, \quad\hbox{and}\quad \cov{R_{t_1,2}, R_{t_2,2}} = \sigma_2^2, \quad t,t_1,t_2\in\mathbb{N}_0.
  \end{equation}

\end{enumerate}

 Models involving both time-invariant and time-varying random effects have previously been considered by \cite{pinquet2020poisson}. Note that here, the process $(R_{t,1}+R_{t,2})$ is stationary but is generically non-Markov.

We can easily check that
\[
\Var{Y_t}=\psi\E{V\left(R_{t,1}+R_{t,2}\right)} +\left( \Var{R_{t,1}}+\Var{R_{t,2}}\right),
\]
and
\[
\cov{Y_{t_1}, Y_{t_2}} = \sigma_1^2\rho^{|t_1-t_2|}+\sigma_2^2.
\]

When $\psi=0$, the covariance matrix of $(Y_1, \cdots, Y_T)$ has the same structure as in \eqref{eq.18}.
Consequently, $\psi\rightarrow 0$, Lemma \ref{lem.3} in \ref{sec.2.3.1} shows the following inequality   
\begin{equation}\label{eq.17}
\widehat{\alpha}_1>\widehat{\alpha}_j, \quad j=2, \cdots, t-1
\end{equation}
which clearly shows that the credibility factors do not satisfy the increasing property in \eqref{eq.16}.

For a numerical illustration, we assume identity variance function $V$ and set  parameters
\[
\sigma_1^2=1,\quad \rho=0.8,
\]
and
\[
\lambda_t=1, \quad t=0, 1, \cdots.
\]
Table \ref{tab.11} shows the credibility factors for $T=5$ for various combinations of $\psi$ and $\sigma_2^2$. As expected, we have
\[
\widehat{\alpha}_1>\widehat{\alpha}_t, \quad t=2, \cdots, 4
\]
under Scenarios I and II with small $\psi=0.01$ and $0.1$, respectively.

\begin{table}[h]
\caption{credibility factors for various combinations of $\psi$ and $\sigma_2^2$} \label{tab.11}
\centering
\begin{tabular}{c|c||c|c|c|c|c|c}
  \hline
  Scenarios & Parameters & $\widehat{\alpha}_1$ & $\widehat{\alpha}_2$ & $\widehat{\alpha}_3$ & $\widehat{\alpha}_4$ & $\widehat{\alpha}_5$ & monotonicity of $(\widehat{\alpha}_j)_j$ \\
  \hline
  I & $\psi=0.01$ and $\sigma_2^2=1$ & 0.046 & 0.011 & 0.011 & 0.042 & 0.805 & no\\
II &  $\psi=0.1$ and $\sigma_2^2=1$ &  0.049 & 0.030 & 0.050 & 0.158 & 0.600 & no\\
III & $\psi=1$ and $\sigma_2^2=1$ & 0.086 & 0.093 & 0.118 & 0.169 & 0.260 & yes \\
IV & $\psi=0.1$ and $\sigma_2^2=0.01$ & 0.003 & 0.009 & 0.034 & 0.137 & 0.554 &yes \\
  \hline
\end{tabular}
\end{table}

However, under Scenarios III and IV, which correspond to
the relatively large value of $\psi=1$, we have the ordering of the credibility factors in \eqref{eq.16}.\footnote{In the case where $\sigma_2^2=0$ with identity variance function $V$,
the covariance matrix of $(Y_1, \cdots, Y_T)$ has the same structure as that of Model \ref{mod.1}. Hence,
Theorem \ref{thm.2} in the later section implies the ordering of the credibility factors in \eqref{eq.16}.  }
\end{example}

\begin{remark}
It is well known that the credibility model is equivalent to regressing $X_t$ on its lagged values. Therefore, the fact that the credibility coefficients, or equivalently the regression coefficients, are not necessarily ordered corresponds to also a well-known issue in the time series literature.\footnote{We thank an anonymous reviewer for this comment.} Let us, for instance, consider
$\hbox{ARMA(1,1)}$ process $(Y_t)$ defined by
\[
Y_t=\phi Y_{t-1}+e_t+\theta-\theta e_{t-1},
\]
where $(e_t)$ represents an unobserved white noise series with zero mean and variance $\sigma_e^2$. Following the standard procedure, we have
\[
\Var{Y_t}=\frac{1-2\phi\theta+\theta^2}{1-\phi^2}\sigma_e^2
\quad\hbox{and}\quad \cov{Y_t, Y_{t+1}}=\phi \Var{Y_t}-\theta\sigma_e^2
\]
and
\[
\cov{Y_t, Y_{t+k}}=\phi\cov{Y_t, Y_{t+k-1}}, \quad \hbox{for}\quad k\ge 2.
\]

Now, consider $\hbox{ARMA(1,1)}$ process $(Y_t)$ with
\[
(\phi, \theta, \sigma_e^2)=(0.5, -0.2, 1.0).
\]
Clearly, $(Y_t)$ has positive covariance ordering. However, the ordering of the credibility factors in \eqref{eq.16} is not satisfied. For example, when $T=5$, the credibility factors in \eqref{eq.11} are given by
\[
(\widehat{\alpha}_1, \cdots, \widehat{\alpha}_5)=(0.001,-0.006, 0.028, -0.140, 0.700),
\]
which shows neither monotonically increasing nor monotonically decreasing pattern.

This issue has (partly) prompted the introduction of alternative time series models that ensure the correct ordering, such as the Harvey-Fernandez model (see Section 3).
\end{remark}



%
%
%
%

\section{Ordering of credibility factors under the AR(1) state-space model}\label{sec.3}

The examples in Section \ref{sec.2.3} show that the monotonicity of the ACF of the state process $(R_t)$ does not guarantee the ordering of the credibility factors corresponding to observable time series $(Y_t)$. In some sense, these counter-examples are not surprising, as \cite{pinquet2020poisson} has already found examples for which positive ACF does not necessarily lead to positive credibility coefficients. In other words, for the credibility coefficients to be ``well behaved,'' generally, further restrictions on the ACF are required.
This section discusses the ordering of credibility factors under the dynamic random effects model and shows that the dynamic random effects model with the random effects having an AR(1)-type ACF ensures the ordering of the credibility coefficients.



\subsection{Model specification and main result}


\begin{model}[AR(1) State-Space Model]\label{mod.1}\footnote{Notably, the whole distributional assumption in Model \ref{mod.1} can be replaced with the moment conditions in \eqref{eq.m1} and \eqref{eq.m2} to give the same result to be presented in this section. However, we stick to the distributional assumption in Model \ref{mod.1} for brevity.}
  Consider a state-space model comprising a state process $(R_t:t=0,1, \ldots)$ and observable time series $(Y_t:t=1,2, \ldots)$ of the following form:
  \begin{enumerate}
    \item[i.] Conditional distribution of $Y_t$ for given $R_t$ is
    \begin{equation}\label{eq.ahn.1}
    Y_t|R_t \sim {\rm ED}\left( \lambda_t R_t, \psi\right)
    \end{equation}
    where ${\rm ED}\left( \lambda_t R_t, \psi\right)$ denotes EDF with mean $\lambda_t R_t$, dispersion parameter $\psi$, and variance function\footnote{From the definition, we have $\E{Y_t|R_t}=\lambda_t R_t$ and $\Var{Y_t|R_t}=\psi V(\lambda_t R_t)$.} $V(\cdot)$.
    \item[ii.] A state process $(R_t:t=0,1, \ldots)$ has stationary Markov property, and we further assume $\E{R_t}=1$ for $t=0,1, \cdots$. Owing to stationary property, $\Var{R_t}$'s are time-invariant constant, and we denote it as
    \begin{equation}\label{eq.241}
   \Var{R_t}=\sigma^2, \quad t=0,1, \cdots
    \end{equation}
    for convenience.
    \item[iii.] We further assume that the ACF satisfies\footnote{Examples of Markov processes $(R_t:t=0,1, \ldots)$ satisfying the correlation structure in \eqref{eq.241} and \eqref{eq.24} will be provided in Section \ref{sec.3.2}.}
    \begin{equation}\label{eq.24}
  \cov{R_{t_1}, R_{t_2}}=\sigma^2\rho^{|t_2-t_1|}, \quad t_1\neq t_2
    \end{equation}
    for $t_1, t_2=0, 1, \cdots$.

  \end{enumerate}
\end{model}

It is straightforward to observe that the closed-form expression of the marginal mean and variance of $Y_t$ is given by
    \begin{equation}\label{eq.m1}
      \E{Y_t}=\lambda_t \quad\hbox{and}\quad\Var{Y_t}=\psi\E{V(\lambda_t R_t)}+\lambda_t^2\sigma^2,
\end{equation}
and the marginal covariance of $Y_{t_1}$ and $Y_{t_2}$ is given by
    \begin{equation}\label{eq.m2}
      \cov{Y_{t_1}, Y_{t_2}}=\lambda_{t_1}\lambda_{t_2} \sigma^2\rho^{|t_1-t_2|}, \quad t_1\neq t_2.
\end{equation}

We first observe that the positive covariance ordering of $(Y_t)$ is not guaranteed depending on the choice of $\lambda_t$'s. To ensure the positive covariance ordering of $(Y_t)$, we restrict our attention to the standardized observations.
Specifically, under the settings in Model \ref{mod.1},
the {\it standardized observations} is defined as
\[
Y_t^*=\frac{Y_t}{\lambda_t}, \quad t=1, 2, \cdots.
\]
In particular, in insurance, standardized observations play a key role in ratemaking systems, and premiums are often expressed in terms of the standardized observations \citep{buhlmann2006course}. For example,
the credibility premium of $Y_{T+1}$ defined in \eqref{eq.10} can be equivalently represented as
\begin{equation}\label{eq51}
{\rm Prem}\left(Y_1, \cdots, Y_T\right):=\widehat{\alpha}_0+\widehat{\alpha}_1^*Y^*_1+\cdots+\widehat{\alpha}_T^*Y^*_T,
\end{equation}
where 
\begin{equation}\label{eq.161}
\widehat{\alpha}_1^*=\lambda_1\widehat{\alpha}_1, \ldots, \widehat{\alpha}_T^*=\lambda_T\widehat{\alpha}_T
\end{equation}
with $\widehat{\alpha}_t$'s given by \eqref{eq.11}.
We say the credibility premium is {\it positively isotonic} if
\begin{equation}\label{eq.160}
\widehat{\alpha}_1^*\le \cdots\le\widehat{\alpha}_T^*.
\end{equation}
We emphasize the terminology ``positive'' here. While we say the sequence has the positive covariance ordering if the ACF is decreasing function in time as in \eqref{eq.2}, we define the credibility premium to be positively ordered if the credibility factors are increasing function in time as in \eqref{eq.160}.

We now have all the necessary ingredients to introduce the main result of the study. The following theorem says that the monotonicity of the credibility coefficients is necessarily satisfied when the dynamic random effect has an AR(1)-type ACF.
\begin{theorem}\label{thm.2}
Consider the time series $(Y_t)$ in the state-space model in Model \ref{mod.1} and
 assume that $\left(\lambda_{1:t}\right)$  is time-invariant:
    \begin{equation}\label{eq.same}
    \lambda_1=\cdots=\lambda_T.
    \end{equation}
 Subsequently, the time series $(Y_t^*)$ has positive covariance ordering, and credibility factors corresponding to the standardized observations are positively isotonic. Specifically,
    $\widehat{\alpha}_1^*,\ldots, \widehat{\alpha}_T^*$ in \eqref{eq51} satisfy the following inequality
        \[
        \widehat{\alpha}_1^*\le \cdots \le \widehat{\alpha}_T^*,
        \]
        where the equality holds if and only if $\rho=0$.
\end{theorem}
\begin{proof}
The proof for the positive covariance ordering of $(Y_t^*)$ is immediate from the following
\[
      \cov{Y_{t_1}^*, Y_{t_2}^*}= \sigma^2\rho^{|t_1-t_2|}, \quad t_1\neq t_2
\]
obtained from \eqref{eq.m2}. For the proof of the remaining part, see Appendix A.
\end{proof}

  We note that the investigation of the ordering of the credibility factors can be less meaningful if \eqref{eq.same} is not satisfied. If we do not assume \eqref{eq.same}, neither $(Y_t)$ nor $(Y_t^*)$ is guaranteed to be stationary so that discussion of the corresponding credibility factors can be void.
Indeed, it will be shown in Section \ref{sec.5.2} through a numerical study that
 the credibility factors change drastically according to the choice of $(\lambda_1, \cdots, \lambda_T)$.
 Consequently, we conclude that the concept of positively isotonic is more sensible with the condition in \eqref{eq.same}.
 In \ref{app.b}, for the completeness of the paper, we also provide results on the ordering of the credibility factors without assuming the condition in \eqref{eq.same}.


\subsection{Parametric examples}\label{sec.3.2}
The class of time series satisfying the correlation structure in \eqref{eq.24},
which is characterized by
\[
\E{R_t|R_{t-1}}= \phi_1 R_{t-1}+\phi_2,
\]
for real numbers $\phi_1$ and $\phi_2$, includes a wide range of AR(1) models (e.g., \citet{grunwald2000theory}). Let us now provide some examples.

\begin{example}[Beta-Gamma autoregressive process of the first-order (BGAR(1))]
  We say that $(R_t:t=0,1, \cdots)$ is a BGAR(1) [see \citet{lewis1989gamma}] with parameter $(\gamma_1, \gamma_2, \rho)\in\Real_+^3$ if it satisfies
 $$
  R_t = B_t R_{t-1}+G_t, \quad t=0, 1, \cdots,
$$
  where $(B_t:t=0,1, \cdots)$ and $(G_t:t=0,1, \cdots)$ are mutually independent sequences of i.i.d. random variables satisfying
$$
   B_t\sim{\rm Beta}\left(\gamma_1\rho, \gamma_1(1-\rho) \right)\quad\hbox{and}\quad G_t\sim {\rm Gamma}\left(\frac{\gamma_1(1-\rho)}{\gamma_2}, \frac{1}{\gamma_1(1-\rho)}\right), \quad t=0,1, \cdots.
 $$
Subsequently, we remark that if $R_{t-1}\sim {\rm Gamma}(\gamma_1/\gamma_2, 1/\gamma_1)$, then by the property of the beta distribution,
we have $$B_t R_{t-1}\sim {\rm Gamma}\left(\frac{\gamma_1\rho}{\gamma_2}, \frac{1}{\gamma_1\rho}\right)$$, and hence, the marginal distribution of $R_t$ is given by
 $$
  R_t \sim {\rm Gamma}\left(\frac{\gamma_1}{\gamma_2}, \frac{1}{\gamma_1}\right).
  $$
Finally, the serial variance and covariance of $(R_t)$ are obtained as
  $$
  \Var{R_t}=\frac{\gamma_1}{\gamma_2^2}\quad\hbox{and}\quad
  \cov{R_t, R_{t-\ell}}=\rho^{t-\ell}\frac{\gamma_1}{\gamma_2^2}, \quad \ell=0, 1, \cdots.
  $$
\end{example}
  The following two examples of state-space model are described by a BGAR(1) with parameter $(\gamma_1, \gamma_2, \rho)$. For the identification purpose, we assume that $\gamma_1=\gamma_2=1/\sigma^2$ so that $\E{R_t}=1$ and $\Var{R_t}=\sigma^2$. Note that we also have $\E{R^3}=(2\sigma^2+1)(\sigma^2+1)$.

  \begin{example}[A dynamic gamma-gamma model]\label{exam.4}
    Consider the state-space model in Model \ref{mod.1} with observable time series $(Y_t: t=1,2, \cdots)$ and
      state variable $(R_t: t=1,2, \cdots)$ with the following specification:
     \begin{itemize}
       \item $(R_t: t=1,2, \cdots)$ is given by a BGAR(1) with parameter $(\gamma, \gamma, \rho)$ with $\gamma=1/\sigma^2$.
       \item The observation $Y_t$ conditional on $R_t$ is given by
    \begin{equation}\label{ex.eq.4}
    Y_t|R_t \sim {\rm Gamma}\left(\lambda_t R_t, \psi \right).
    \end{equation}
     \end{itemize}
%
    Subsequently, for $t, t_1, t_2\in \mathbb{N}$ and $t_1\neq t_2$, we have
    \[
    \E{Y_t}=\lambda_t, \quad \Var{Y_t}=\lambda_t^2\left(\psi\left( 1+\sigma^2\right)+\sigma^2 \right), \quad \hbox{and}\quad
    \cov{Y_{t_1}, Y_{t_2}}=\lambda_{t_1}\lambda_{t_2}\rho^{|t_1-t_2|}\sigma^2.
    \]
    Furthermore, using Theorem \ref{thm.1} in \ref{app.a}, $\widehat{\alpha}_1^*,\ldots, \widehat{\alpha}_T^*$ in \eqref{eq51} is obtained as
    \begin{equation}\label{buh.2}
    \left(\widehat{\alpha}_1^*,\ldots, \widehat{\alpha}_T^*  \right)=
        \rho\left(1-\rho^2\right)\sigma^2\lambda_{T+1} v_T
        \left(u_1, \ldots, u_T \right)\frac{1}{\left(1+\sigma^2\right)\psi},
    \end{equation}
    where $u_1, \ldots, u_T$ and $v_T$ are defined in Definition \ref{def.2} in \ref{app.a}.
    Clearly, the credibility premium is regular. Furthermore, if $\lambda_1=\cdots=\lambda_T$, the time series $(Y_t)$ has positive covariance ordering.
    Finally, with the help of Lemma \ref{lem.1}, the results in \eqref{buh.2} clearly show that the credibility premium is regular. This confirms the result in Theorem \ref{thm.2}.


  \end{example}

  \begin{example}[A dynamic Poisson-gamma model]\label{exam.3}
    Consider the same state-space model in Model \ref{mod.1}, where the distributional assumption in \eqref{ex.eq.4} is replaced by
    \begin{equation}\label{ex.eq.5}
    Y_t|R_t \sim {\rm Pois}\left(\lambda_t R_t \right).
    \end{equation}
    Subsequently, for $t, t_1, t_2\in \mathbb{N}$ and $t_1\neq t_2$, we have
    \[
    \E{Y_t}=\lambda_t, \quad \Var{Y_t}=\lambda_t+\lambda_t^2\sigma^2, \quad \hbox{and}\quad
    \cov{Y_{t_1}, Y_{t_2}}=\lambda_{t_1}\lambda_{t_2}\sigma^2 \rho^{|t_1-t_2|}.
    \]
        Similar to Example \ref{exam.4}, Theorem \ref{thm.1} in Appendix A shows that $\widehat{\alpha}_1^*,\ldots, \widehat{\alpha}_T^*$ in \eqref{eq51} are obtained as
    \[
    \left(\widehat{\alpha}_1^*,\ldots, \widehat{\alpha}_T^*  \right)=
        \rho\left(1-\rho^2\right)\sigma^2\lambda_{T+1} v_T \left(u_1\lambda_1, \ldots, u_T\lambda_T \right).
    \]
    where $u_1, \ldots, u_T$, and $v_T$ are defined in Definition \ref{def.2} in Appendix A.
    Clearly, the credibility premium is regular. Furthermore, if $\lambda_1=\cdots=\lambda_T$, the time series $(Y_t)$ has positive covariance ordering, and credibility premium is positively isotonic by Lemma \ref{lem.1}. This confirms the result in Theorem \ref{thm.2}.
  \end{example}


\begin{remark}\label{rem.1}
Under certain circumstances, the credibility premium is guaranteed to be positively isotonic without assuming time-invariant $(\lambda_t)$ as in \eqref{eq.same}.
Let us, for instance, reconsider Example \ref{exam.4}.
As the unit variance function of Gamma distribution is $V(x)=x^2$, part ii of Corollary \ref{app.thm.1} in \ref{app.b} shows that the credibility premium is guaranteed to be positively isotonic without the assumption in \eqref{eq.same}.

Similar ideas can be used to come up with the example whose (non-standard) credibility factors are positively ordered without assuming \eqref{eq.same}. Consider the time series $(Y_t)$  in \eqref{exam.3}.
As the unit variance function of Poisson distribution is the identity function---$V(x)=x$---part i of Corollary \ref{app.thm.1} in \ref{app.b} shows that the credibility factors satisfy
        \[
        \widehat{\alpha}_1\le \cdots \le \widehat{\alpha}_T
        \]
regardless of the assumption in \eqref{eq.same}.

However, as we have already discussed, neither the time series $(y_t)$ nor $(y_t^*)$ is stationary. Consequently, it is less meaningful to discuss the ordering of credibility factors without assuming time-invariant $(\lambda_t)$ as in \eqref{eq.same}.


\end{remark}

%

Finally, note that besides BGAR(1) used in Examples \ref{exam.4} and \ref{exam.3}, there are other stationary processes with stationary, gamma distribution, as well as AR(1)-type auto-covariance function. In the following, let us give two well-known examples.

\begin{example}[Autoregressive gamma process (ARG)]
	The ARG process is the exact time-discretization of the square-root diffusion process and has been introduced into the ratemaking literature by \cite{lu2018dynamic}. This process is Markov and is mostly conveniently specified through the conditional Laplace transform:
\begin{equation}
\label{lt}
\mathbb{E}[\exp(-s R_{t+1})\mid R_t] =\frac{1}{(1+cs)^{\delta}} \exp\Big(-\frac{\rho s}{1+cs}R_t\Big), \qquad \forall s \geq 0,
\end{equation}
where $\rho \in [0,1]$, and $c, \delta>0$. By taking first derivative with respect to $s$, it is easily shown that $\mathbb{E}[R_{t+1}|R_t]=\rho R_t+c\delta$.

As this model has the same auto-covariance function as the above BGAR(1) process, it gives rise to the same credibility coefficients in a state-space context.\footnote{See also \cite{lu2018dynamic} for the expression of Bayes premium when the response variable is conditionally Poisson and \cite{li2020dynamic} for numerical approximation methods in the general case. }
\end{example}

\begin{example}[Gamma autoregressive process]
\cite{gaver1980first} consider the process $(R_t)$ defined by
$$R_{t+1}=\rho R_t+\epsilon_{t+1},$$
where the innovation process $(\epsilon_t)$ is i.i.d. They work out the distribution of the latter such that process $(R_t)$ is marginally gamma distributed. For this example, by definition, we have $\mathbb{E}[R_{t+1}|R_t]=\rho R_t+\mathbb{E}[\epsilon_t]$.
\end{example}

\subsection{Comparison with Harvey and Fernandez's approach}
The dynamic, AR(1)-type random effects model considered above is a state-space model with an exogenous state process. In other words, the conditional distribution of $R_t$ given the past of both processes $(R_t)$ and $(y_t)$ only depends on its own past but not on that of $(y_t)$. Alternatively, the time series literature has also proposed state-space models with non-exogenous or endogenous state process, which does not have the causal chain \eqref{chainstructure}. One such example is the count time series model of \cite{harvey1989time}, which has been applied in the actuarial literature by \cite{bolance2007greatest, abdallah2016sarmanov}. This model, which is based on the Poisson-gamma conjugacy,  is such that at any time $t$, both the filtering density $R_t| (Y_1, \cdots, Y_t)$ and the predictive density $R_{t+1}| (Y_1, \cdots, Y_t)$ are gamma. More precisely, the joint dynamics of $(R_t, y_t:t=0,1, \ldots)$ is defined as follows:
\begin{itemize}
  \item (Initial value) $R_1 \sim {\rm Gamma}^*(a_0, b_0)$ with $a_0=b_0$.
  \item (measurement equation) For each $t$, \[
  Y_t|(R_t, R_{t-1},..., Y_{t-1},...) \sim {\rm Pois}\left( \lambda_t R_t\right),
  \]
  \item (Updating formula) If
  \[
  R_t| (Y_1, \cdots, Y_t)=(y_1, \cdots, y_t)\sim {\rm Gamma}^*(a_t, b_t),
   \]
  then the one-step-ahead conditional distribution of the random effect is given by
     \begin{equation}
     \label{harvey}
      R_{t+1}| (Y_1, \cdots, Y_t)=(y_1, \cdots, y_t)\sim {\rm Gamma}^*(\alpha a_t, \alpha b_t)
      \end{equation}
      for a time-invariant constant $\alpha\in (0,1]$. Moreover, these authors show that $a_t$ and $b_t$ are some affine functions of $(y_1, \cdots, y_t)$.
\end{itemize}
Here, ${\rm Gamma}^*(a_0, b_0)$ denotes the gamma distribution with shape parameter $a_0$ and rate parameter $b_0$.

A direct consequence of this specification, and especially equation \eqref{harvey}, is the following exponential moving average predictive mean:
\begin{equation}\label{eq.y.1}
\E{Y_{t+1}| Y_1, \cdots, Y_t}=\lambda_{t+1}\frac{a_0+\sum\limits_{\tau=1}^{t} \alpha^{-\tau} Y_{t}}{a_0+\sum\limits_{\tau=1}^{t}\alpha^{-\tau}\lambda_t}.
\end{equation}
As the conditional expectation is linear, the credibility forecast and the conditional expectation coincide. Therefore, this is yet another model in which the credibility coefficients form a decreasing (indeed geometric) sequence.

The major differences between Model \ref{mod.1} and the model of Harvey and Fernandez are three-fold. First, unlike Model \ref{mod.1}, \cite{harvey1989time}'s model is specific to the models with conditional Poisson distribution.\footnote{\cite{smith1986non} proposes a similar model for exponential observations and Gamma random effects, but the latter has not yet been applied to the insurance literature.}
Second and more importantly, the dynamics of the random effects process $(R_t:t=0,1, \ldots)$ defined in \cite{harvey1989time} is nonstationary, in the sense that the marginal distribution of $R_t$ is not time-invariant., even when $\lambda_t$ is time-variant.
This non-stationarity makes the interpretation of this model considerably unnatural. Finally, recently, \cite{boucher2018claim} showed that, unlike random effects type models, the Harvey-Fernandez model has an undesirable numerical feature that the impact of a single claim for an insured is excessively large for policyholders with a long driving experience. This makes it unappealing when it comes to retaining loyal and good drivers.

\section{Numerical study} \label{sec:num}
This section performs numerical studies on the credibility factors with the proposed methodology. First, in Section \ref{sec.5.2}, we provide a numerical example showing that the state-space model in Model \ref{mod.1} is not necessarily positively isotonic. Subsequently, in Section \ref{sec.5.3}, we analyze credibility premiums with the proposed model under various parameter settings in Model \ref{mod.1} and compare their performances to those of the exact premium in (\ref{exact.1}).

\subsection{Numerical examples of credibility premiums which are not positively isotonic}\label{sec.5.2}
In Section \ref{sec.2.3}, we show that the credibility premium under a state-space model is not necessarily positively isotonic, although it has the decreasing covariance property.
In this subsection, we show, by example, that even a state-space model as in Model \ref{mod.1} can have credibility premiums that are not positively isotonic. While Theorem \ref{thm.2} shows that credibility factors under the assumptions in Model \ref{mod.1} are positively isotonic if time-invariant fixed effects are assumed, we may have credibility factors that are not positively isotonic with time-varying fixed effects.

First, consider the credibility premium under the settings in Example \ref{exam.3} with $t=5$,  $\sigma^2=0.5$, and $\psi=0.5$.
For other parameters, consider the combination of $\rho=0.3,0.6$ and
\[
\left(\lambda_1, \cdots, \lambda_5\right)=
\begin{cases}
  (1, 1, 1, 1, 1);\\
  (0.001, 0.01, 0.1, 1, 10);\\
  (10, 1, 0.1, 0.01, 0.001);\\
\end{cases}
\]
with $\lambda_6=1$. Table \ref{tab.1} shows the credibility factors calculated in \eqref{buh.2}.

Columns corresponding to Case 1.a and Case 2.a in Table \ref{tab.1} represent  time-invariant fixed effects.
Clearly, credibility premiums in these cases are positively isotonic, which confirms Theorem \ref{thm.2}.
However, by Case 1.b and 1.c, we show that the ordering of credibility factors can be affected by controlling the values of $(\lambda_1, \cdots, \lambda_T)$. Specifically, while the credibility premium in Case 1.b is positively isotonic, the credibility premium in Case 1.c is not positively isotonic. This example suggests that the fair comparison of the ordering of credibility factors should be based on the assumption
\[
\lambda_1= \cdots =\lambda_T.
\]
We can observe a similar pattern of credibility factors in Case 2.b and 2.c.

\begin{table}[h!]
\caption{Credibility factors of the standardized observations in Example \ref{exam.3}
  (The credibility factors are written in the unit of $0.001$.)} \label{tab.1}
\centering
\begin{tabular}{|c|c|c|c c c c c|}
  \hline
    & $\rho$ & $(\lambda_1, \cdots, \lambda_5)$ & $\widehat{\alpha}_1^*$ & $\widehat{\alpha}_2^*$ & $\widehat{\alpha}_3^*$ & $\widehat{\alpha}_4^*$ & $\widehat{\alpha}_5^*$\\
    \hline
    Case 1.a & 0.3 & $(1, 1, 1, 1, 1)$ & 0.167  &   0.809  &   3.999  &  19.785  &  97.894  \\
    Case 1.b & 0.3 & $(0.001, 0.01, 0.1, 1, 10)$ & 0.000  &   0.004  &   0.147  &   5.114  & 248.710   \\
    Case 1.c & 0.3 & $(10, 1, 0.1, 0.01, 0.001)$ & 1.314  &   2.430   &  1.238  &   0.444  &   0.150   \\
    \hline
    Case 2.a & 0.6 & $(1, 1, 1, 1, 1)$ & 6.172  &  13.578  &  31.847   & 75.594  & 179.815    \\
    Case 2.b & 0.6 & $(0.001, 0.01, 0.1, 1, 10)$ &  0.005  &   0.076   &  1.279  &  22.016  & 488.594  \\
    Case 2.c & 0.6 & $(10, 1, 0.1, 0.01, 0.001)$ & 45.860  &  32.102   &  8.530   &  1.658  &   0.291 \\
  \hline
\end{tabular}
\end{table}

\subsection{Calculation of credibility premiums under various parameter settings}\label{sec.5.3}
We investigate the accuracy and applicability of the proposed credibility premium compared to the exact premium in \eqref{exact.1} and some industry benchmarks.
Here, \(T=5\) and \(Y_{it}\) for
\(t=1,\ldots, 6\) and \(i=1,\ldots, 500\) are generated as follows: \[
Y_{it}|R_{it} \sim {\rm Pois}\left(\lambda_{it} R_{it} \right), \quad \lambda_{it} =\exp(-3+2X_{it}),
\] where \(X_{it}\) follows a normal distribution with mean 0 and
variance 0.6, and $(R_t: 0,1, \cdots)$ is BGAR$(1)$ with parameters $(\gamma, \gamma, \rho)$ with $\gamma=1/\sigma^2$ so that $\E{R_t}=1$ and $\Var{R_t}=\sigma^2$.
With this scheme, we generate eleven hypothetical datasets by varying the values of
\(\sigma^2=0, 1,2\) and \(\rho=0.0, 0.3, 0.6, 0.9, 1.0 \) to capture various scenarios of serial dependence within the claims of the same policyholder. For example, if $\rho=\sigma^2=0$, $Y_{it}$ marginally follows an independent Poisson distribution. If $\rho=0$ and $\sigma^2>0$, then the $Y_{it}$ marginally follows a negative binomial distribution, while there is no serial dependence among $Y_{i,1}, \ldots, Y_{i,6}$ for each $i$. If $\rho=1$, then $Y_{it}$ follows a usual static gamma random effects model explained below. While \(Y_{it}\) for \(t=1,\ldots, 5\) are used as a training set, \(Y_{i,6}\) is set aside as an out-of-sample validation set to assess the predictive performance of each premium.

%
%
%

For the numerical comparison, we consider the following ways of premium calculation as benchmarks under the framework in Model \ref{mod.1}:
\begin{itemize}
  \item \textbf{Naive}: premium without a posteriori ratemaking so that $\hat{\mu}_{i,6}= \hat{\lambda}_{i,6}$.  Note that it is the true premium when $\rho=\sigma^2=0$ in the BGAR(1) model.
  \item \textbf{Static}: premium when, in the BGAR(1) model, the correlation $\rho=1$---when $R_{i,1} = \cdots = R_{i,6}$ for all $i$. In other words, in this case, we have the usual static gamma random effects model, and standard conjugacy leads to
\[
\hat{\mu}_{i,6}= \frac{\sum_{t=1}^5 Y_{it} +1/\sigma^2}{\sum_{t=1}^5 \hat{\lambda}_{it} +1/\sigma^2} \hat{\lambda}_{i,6}.
\]
Here, we estimated \( 1/\sigma^2\) using the method of moments. Alternatively, one can also use \texttt{Hglm} package in \texttt{R} \citep{ronnegaard2010hglm} to find the MLE of $\sigma^2$ or the argument of prior elicitation \citep{jeong2020bregman}.
  \item \textbf{Exact}: premium using Markov Chain Monte Carlo (MCMC) simulations\footnote{We note that the analytical expression of the exact premium in \eqref{exact.1} under the setting in this subsection is not possible. Alternatively, the exact premium in \eqref{exact.1} can be obtained by MCMC simulations.} to sample $R_{i,6}$ for $i=1,\ldots, 500$ so that
$$
\hat{\mu}_{i,6}=  \frac{1}{S}\sum_{s=1}^S \hat{R}^{(s)}_{i,6} \hat{\lambda}_{i,6}.
$$
where $\hat{R}^{(s)}_{i,6}$, $s=1,\ldots, S$ for each $i$ are posterior samples of $R_{i,6}$. 3,000 posterior samples were used for each $R_{i,6}$ so that $S=3,000$. Note that the use of more posterior samples may lead to better prediction results. However, the computation cost is linearly proportional to the number of posterior samples. Consequently, it might not be desirable to extract excessively many posterior samples to accomplish marginal improvement in prediction.

  \item \textbf{True}: premium that uses the true value of $R_{i,6}$ that have been used for generating $Y_{i,6}$ so that $\hat{\mu}_{i,6}= R_{i,6}\hat{\lambda}_{i,6}$. Note that true premium is only available with simulation studies and contains no model error on the unobserved heterogeneity component.
\end{itemize}

We compare the actual claim frequency in the out-of-sample validation set with the predicted claim frequency
using the premium calculation methods for each policyholder. Prediction performances are
measured by mean absolute error (MAE) and root-mean-square error (RMSE),
which quantify discrepancies between the actual and predicted
values using \(L_1\) and \(L_2\) norms, respectively.  To incorporate the inherent randomness of the Poisson random variable $Y_{i,6}$, we generate 100 copies of $Y_{i,6}$ with the fixed mean $R_{i,6}\lambda_{i,6}$ and compute both RMSE and MAE. In Table \ref{tab:valsim}, notably, both the naive and static premiums cannot capture the time-decaying effects of dynamic random effects and have relatively poorer performances when $\rho=0.3, 0.6$, and $0.9$. While the exact premium tends to have better prediction performance, such performance of the exact premium is at the expense of substantial calculation time in this particular example\footnote{All the premiums were calculated using a laptop with Intel Core i7-8565K at 1.80Ghz 4 cores, 16GB memory.} as shown in Table \ref{tab:timesim}. Conversely, the use of the proposed premium is acceptable throughout diverse scenarios of serial dependence, while the computation cost is modest compared to the exact method via MCMC. In this regard, our proposed method can be considered as a reasonable and applicable approximation of the true premium in practice in various scenarios.

\begin{table}[!h]
\caption{\label{tab:}\label{tab:valsim}Relative prediction errors of different premiums using simulated datasets (in percentage)}
\centering
\begin{tabular}[t]{rrrrrrrrrrrr}
\toprule
\multicolumn{2}{c}{ } & \multicolumn{5}{c}{RMSE} & \multicolumn{5}{c}{MAE} \\
\cmidrule(l{3pt}r{3pt}){3-7} \cmidrule(l{3pt}r{3pt}){8-12}
$\rho$ & $\sigma^2$ & Naive & Static & Proposed & Exact & True & Naive & Static &  Proposed & Exact & True\\
\midrule
0.0 & 0 & 100 & 134 & 100 & 100 & 100 & 100 & 105 & 100 & 100 & 100\\
\hline
0.0 & 1 & 121 & 120 & 121 & 121 & 100 & 120 & 120 & 120 & 121 & 100\\
0.3 & 1 & 118 & 118 & 118 & 118 & 100 & 117 & 116 & 116 & 120 & 100\\
0.6 & 1 & 114 & 113 & 111 & 110 & 100 & 124 & 123 & 122 & 117 & 100\\
0.9 & 1 & 114 & 112 & 112 & 111 & 100 & 119 & 118 & 116 & 113 & 100\\
1.0 & 1 & 113 & 109 & 109 & 110 & 100 & 116 & 113 & 110 & 112 & 100\\
\hline
0.0 & 2 & 132 & 132 & 132 & 132 & 100 & 133 & 132 & 133 & 134 & 100\\
0.3 & 2 & 117 & 117 & 117 & 117 & 100 & 126 & 126 & 125 & 128 & 100\\
0.6 & 2 & 117 & 132 & 115 & 118 & 100 & 132 & 137 & 131 & 137 & 100\\
0.9 & 2 & 118 & 116 & 113 & 115 & 100 & 129 & 127 & 123 & 117 & 100\\
1.0 & 2 & 126 & 121 & 117 & 123 & 100 & 135 & 128 & 119 & 124 & 100\\
\hhline{============}
\end{tabular}
\end{table}

\begin{table}[!h]
\caption{\label{tab:}\label{tab:timesim}Computation times of different premiums using simulated datasets}
\centering
\begin{tabular}[t]{rrrrrr}
\toprule
$\rho$ & $\sigma^2$ & Naive & Static &  Proposed & Exact\\
\midrule
0.0 & 0 & 0.04 & 0.05 & 2.75 & 792.08\\
\hline
0.0 & 1 & 0.04 & 0.05 & 2.88 & 1360.26\\
0.3 & 1 & 0.04 & 0.05 & 2.91 & 995.59\\
0.6 & 1 & 0.04 & 0.06 & 2.86 & 739.62\\
0.9 & 1 & 0.04 & 0.05 & 2.93 & 786.19\\
1.0 & 1 & 0.04 & 0.05 & 3.10 & 814.42\\
\hline
0.0 & 2 & 0.04 & 0.05 & 3.01 & 1329.40\\
0.3 & 2 & 0.04 & 0.05 & 3.01 & 1394.84\\
0.6 & 2 & 0.04 & 0.06 & 2.93 & 1419.82\\
0.9 & 2 & 0.04 & 0.05 & 2.96 & 766.59\\
1.0 & 2 & 0.04 & 0.05 & 2.91 & 736.12\\
\hhline{======}
\end{tabular}
\end{table}


\pagebreak

\section{Actuarial application: ratemaking with longitudinal data}

For a real data analysis, we use a sample of claim dataset from the
Local Government Property Insurance Fund (LGPIF), operated by the state
of Wisconsin. Although this dataset is considerably rich and contains information on multiple lines of insurance coverage, here, we only focus on the posterior ratemaking of inland marine (IM) claims with our proposed
state-space model. This sample contains 6,338 observations of claims and
policy characteristics over years 2006--2011 with 1,098 policyholders.
Observations over the years 2006--2010 are used as the training set, while
observations over the year 2011 are set aside for out-of-sample validation.
Table \ref{tab:datasummary} provides brief summary statistics of policy
characteristics. There are seven categorical explanatory variables
related with location and two continuous covariates. We direct
interested readers to refer to \citet{frees2016multivariate} for a more detailed preliminary analysis of the dataset.

\begin{table}[h!t!]
\begin{center}
\caption{Observable policy characteristics used as covariates} \label{tab:datasummary}
\resizebox{!}{3cm}{
\begin{tabular}{l|lrrr}
\hline \hline
Categorical & Description &  & \multicolumn{2}{c}{Proportions} \\
variables \\
\hline
TypeCity & Indicator for city entity:           & Y=1 & \multicolumn{2}{c}{14.50 \%} \\
TypeCounty & Indicator for county entity:       & Y=1 & \multicolumn{2}{c}{5.92 \%} \\
TypeMisc & Indicator for miscellaneous entity:  & Y=1 & \multicolumn{2}{c}{10.78 \%} \\
TypeSchool & Indicator for school entity:       & Y=1 & \multicolumn{2}{c}{29.10 \%} \\
TypeTown & Indicator for town entity:           & Y=1 & \multicolumn{2}{c}{16.60 \%} \\
TypeVillage & Indicator for village entity:     & Y=1 & \multicolumn{2}{c}{23.09 \%} \\
NoClaimCreditIM & No IM claim {in three consecutive prior years}:    & Y=1 & \multicolumn{2}{c}{43.99 \%} \\
\hline
 Continuous & & Minimum & Mean & Maximum \\
 variables \\
\hline
CoverageIM  & Log coverage amount of IM claim in mm  &  0 & 0.87
            & 46.75\\
lnDeductIM  & Log deductible amount for IM claim     &  0 & 7.14
            & 9.21\\
\hline \hline
\end{tabular}}
\end{center}
\end{table}

Table \ref{tab:betasum} provides estimated \(\beta\) and corresponding
standard errors. Similar to the simulation study in Section \ref{sec.5.3}, we commonly use $\hat{\beta}$ estimated from usual Poisson GLM for all the premiums to focus on the impact of each methodology on posterior ratemaking.
\begin{table}[!h]

\caption{\label{tab:unnamed-chunk-3}\label{tab:betasum}Regression estimates of the frequency models}
\centering
\begin{tabular}[t]{lrrr}
\toprule
  & Estimate & Std. err & p-value\\
\midrule
(Intercept) & -4.0315 & 0.3204 & 0.0000\\
TypeCity & 0.9437 & 0.1907 & 0.0000\\
TypeCounty & 1.7300 & 0.1993 & 0.0000\\
TypeMisc & -2.7326 & 1.0149 & 0.0071\\
TypeSchool & -0.9172 & 0.2776 & 0.0010\\
TypeTown & -0.3960 & 0.2772 & 0.1531\\
CoverageIM & 0.0664 & 0.0074 & 0.0000\\
lnDeductIM & 0.1353 & 0.0458 & 0.0031\\
NoClaimCreditIM & -0.3690 & 0.1313 & 0.0050\\
\hhline{====}
\end{tabular}
\end{table}

To compute credibility factors in the proposed model, we need to
determine the values of \(\rho\) and \(\gamma\). We can also
consistently estimate dispersion parameters of the state-space model,
\(\rho\) and \(\gamma\), using the method of moments---a similar
approach to page 13 of \citet{sutradhar2003mixedpoisson}. It turns out
that \(\hat{\rho}=0.8831\) and \(\hat{\gamma}=0.8360\).

\begin{figure}
\centering
\includegraphics{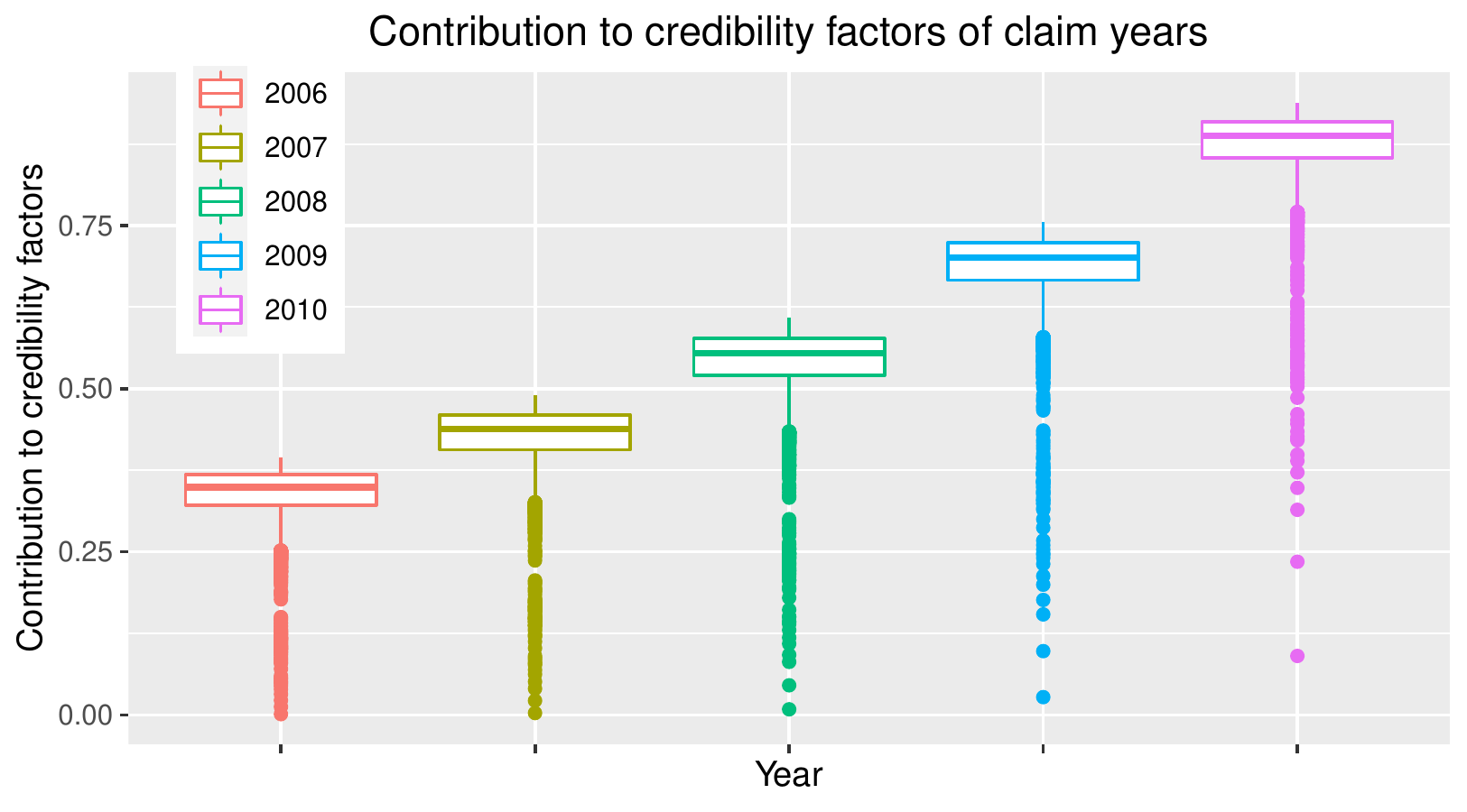}
\caption{\label{fig:credcont}Contribution to credibility factors of past
claim frequencies}
\end{figure}

Once \(\hat{\beta}\), \(\hat{\rho}\), and \(\hat{\gamma}\) are
determined, one can compute credibility factors for the proposed model based on \eqref{eq.22} and \eqref{eq.33}. We expect that
\(\widehat{\alpha}_1 \leq \cdots \leq \widehat{\alpha}_5\)
from Remark \ref{rem.1}. Figure \ref{fig:credcont}
visualizes contributions to credibility factors,
\(\left(\widehat{\alpha}_1/\lambda_{T+1}, \,\ldots, \, \widehat{\alpha}_T/\lambda_{T+1} \right)\),
along with the ages of past claims that also show clearly increasing
patterns of credibility factors as well as the nonnegativity of the credibility factors.

Since
\(\widehat{\alpha}_{0}^*= \lambda_{T+1} - \sum_{t=1}^T \widehat{\alpha}_t \lambda_t = \lambda_{T+1} - \sum_{t=1}^T \widehat{\alpha}^*_t\),
we have the following:

\begin{equation} \label{eqn:cred}
\begin{aligned}
{\rm Prem}\left(Y_1, \cdots, Y_T\right) &= \widehat{\alpha}^*_0 + \sum_{t=1}^T \widehat{\alpha}_t^* Y_t^* = \lambda_{T+1} + \sum_{t=1}^T \widehat{\alpha}_t^*(Y_t^*-1) \\
&=\lambda_{T+1} + \sum_{t=1}^T \widehat{\alpha}_t(Y_t-\lambda_t) = \lambda_{T+1} \left( 1+\sum_{t=1}^T   \frac{\widehat{\alpha}_t}{\lambda_{T+1}}(Y_t-\lambda_t) \right).\\
\end{aligned}
\end{equation} Therefore, one can interpret
\(1+\sum_{t=1}^T \frac{\widehat{\alpha}_t}{\lambda_{T+1}}(Y_t-\lambda_t)\)
as a posterior rating factor, which is multiplied by the prior premium to
reflect past claims history with adjustments.

\begin{figure}
\centering
\includegraphics{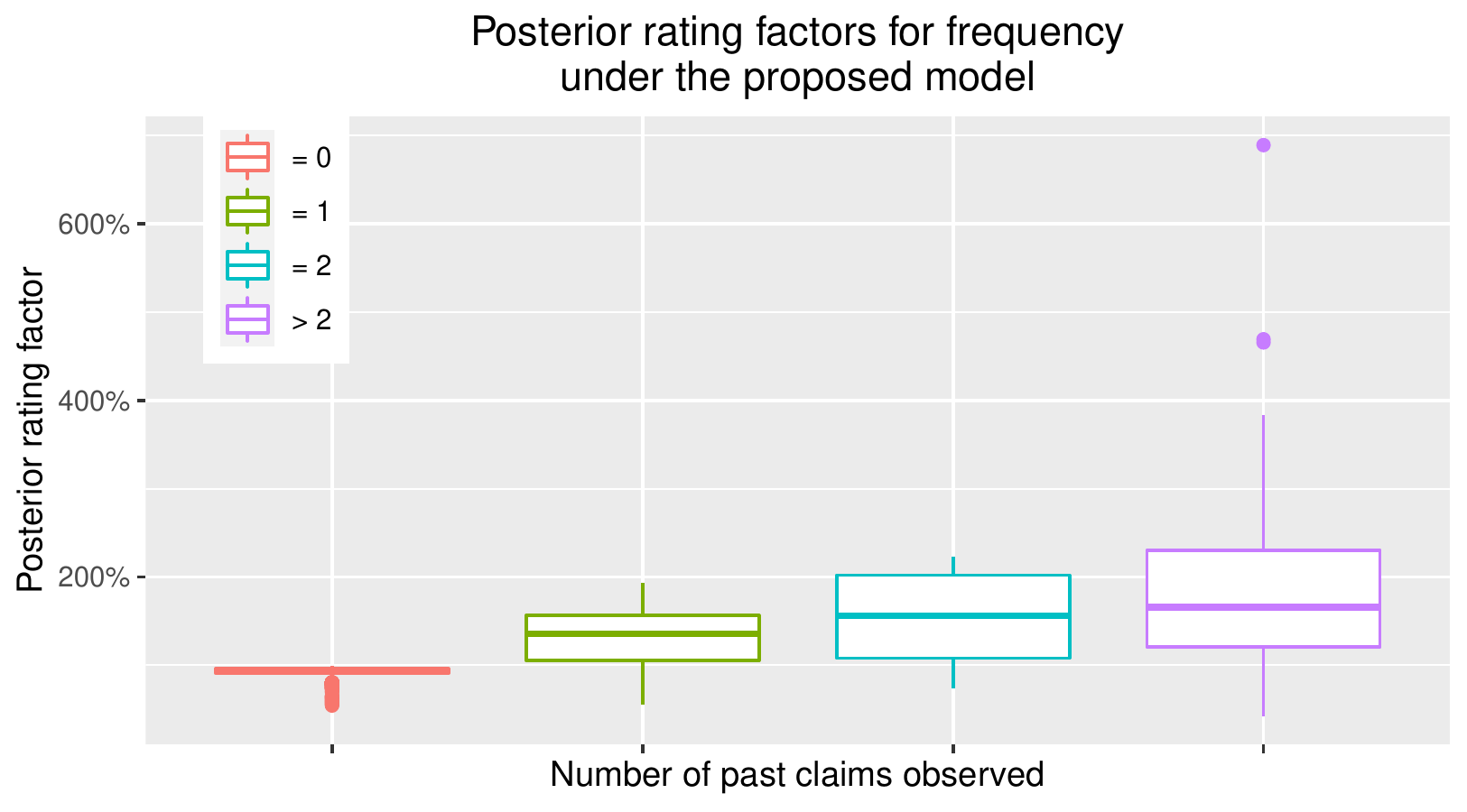}
\caption{\label{fig:credpast}Distribution of posterior rating factor
with past claim frequencies}
\end{figure}

Figure \ref{fig:credpast} explains the behaviors of posterior rating
factors depending on the number of past claims observed. First, those who
had no claim at all in the past years would get sure discounts, which is
considerably reasonable and agree with (\ref{eqn:cred}) as
\(Y_t-\lambda_t<0\) for all \(t=1,\ldots, T\) in this case. One can also
see that posterior rating factors tend to increase as a function of
the numbers of past claims on average, which gives a motivation of risk
mitigation and control to the policyholders for possible future
discount. Further, interestingly, it is still possible to get
discount even with a positive number of past claims as
\(Y_t-\lambda_t\) is not necessarily positive, especially when
\(\lambda_t\) is sufficiently large. Therefore, the given posterior ratemaking scheme still penalizes those who had claims in the past but in a
relative manner and considering adjustments with varying levels of
expected claims determined by observable characteristics.

In this sample study, the proposed dynamic credibility premium outperforms both the naive and static credibility premiums as shown in Table \ref{tab:valtable}. This is in terms of both RMSE and MAE and even as good as the exact premium, which is obtained by 30,000 posterior sampling of the posterior rating factor via MCMC with substantial computation time reduction.\footnote{In this small sample study, the proposed dynamic credibility premium ourperforms the exact premium in the test sample.}

\begin{table}[!h]

\caption{\label{tab:unnamed-chunk-5}\label{tab:valtable}Validation measures for the frequency models}
\centering
\begin{tabular}[t]{lrrrr}
\hhline{=====}
  & Naive & Static & Proposed & Exact \\
\midrule
RMSE & 0.6439 & 0.5002 & 0.4263 & 0.4661 \\
MAE & 0.1220 & 0.1121 & 0.1046 & 0.1060\\
\hhline{=====}
\end{tabular}
\end{table}

\section{Concluding remarks}
In this study, we have examined the existence or the lack thereof of decreasing properties for the credibility coefficients in dynamic random effect models. We have shown that this is the case when the dynamic random effect process has an AR(1) type ACF and have found counter-examples when the ACF assumes some of the more general forms. Hence, for these models, credibility premium can be safely used, if the reader is unwilling to specify completely the dynamics of the random effect process for computational or robustness concerns.
However, because the AR(1)-type ACF is a quite restrictive specification,
the take-home message we want to convey is \textit{not necessarily} to stick with AR(1)-type dynamic credibility models. Rather, we also want to bring this \textit{potential} limitation of dynamic credibility to the attention of the actuarial community. In particular, one important open question is whether we can suitably extend the class of ``admissible'' ACF's beyond the AR(1) case. Indeed, in the two counter-examples that we found, the dynamic random effect process has an unconstrained ACF, or is non-ergodic.
In other words, we have not ruled out the possibility that other ergodic processes with non-AR(1) ACF could be compatible with decreasing credibility coefficients. For instance, some interesting candidate ACF's to investigate would be those of ARFIMA(0,$d$,0)-type, which have recently been proved by \cite{pinquet2020positivity} to ensure at least the positivity of the credibilities. This is left for future research.

\section*{Acknowledgements}

Jae Youn Ahn was supported by a National Research Foundation of Korea (NRF) grant funded by the Korean Government (2020R1F1A1A01061202). Himchan Jeong was supported by the Simon Fraser University New Faculty Start-up Grant (NFSG). Yang Lu thanks CNRS (France) for a teaching release grant and Concordia University for a Start-up grant as well as NSERC through a discovery grant (RGPIN-2021-04144, DGECR-2021-00330). The authors wish to thank Prof. Jean Pinquet and Prof. Dong Wan Shin for their comments.

\bibliographystyle{plainnat}
\bibliography{CTE_Bib_HIX}

\appendix
\section{Proof of Theorem \ref{thm.2}}\label{app.a}

For expository purpose, let us first introduce some further notations. Define $T\times T $ diagonal matrices
\[
\boldsymbol{\Lambda}_T={\rm diag}\left(\lambda_1, \cdots, \lambda_T\right)
\quad\hbox{and}\quad
\boldsymbol{\Xi}_T={\rm diag}\left(\xi_1, \cdots, \xi_T\right),
\]
where
\[
\xi_t:=\frac{\sigma^2\left( 1-\rho^2\right)}{\psi}
\frac{\left(\lambda_t \right)^2}{\E{V\left(\lambda_tR_t\right)}}, \quad t=1, \ldots, T,
\]
and define the $T \times T$ covariance matrix $\boldsymbol{\Sigma}_T$ of $\left(Y_{1:T}\right)$ as
\[
\left[\boldsymbol{\Sigma}_T\right]_{t_1,t_2}=
\begin{cases}
  \cov{Y_{t_1}, Y_{t_2} }, & t_1\neq t_2;\\
  \Var{Y_{t_1}}, & t_1=t_2;
\end{cases}
\]
for $t_1, t_2=1, \cdots, T$.
Further, we define the following symbols, which are convenient to express the result on credibility premium for Model \ref{mod.1}.

\begin{definition}\label{def.2}
For the given $\xi_1, \ldots, \xi_T>0$ and $\rho\in(-1,1)$, define the sequences $\left(v_1, \ldots, v_T\right)$ and $\left(u_1, \ldots, u_T\right)$ as follows.
  \begin{enumerate}
    \item[i.] Define $d_T:=1+\xi_T$ and iteratively define
    \[
    d_t:=1+\rho^2+\xi_t-\frac{\rho^2}{d_{t+1}}, \quad \hbox{for}\quad
    t=T-1, \ldots, 2.
    \]
    Finally, define
    \[
    d_1:=1+\xi_1-\frac{\rho^2}{d_2}.
    \]
    \item[ii.] Define $v_1=1/d_1$ and iteratively define
    \[
    v_t:=\frac{\rho}{d_t}v_{t-1}, \quad \hbox{for}\quad
    t=2, \ldots, T.
    \]
    \item[iii.] Define $\delta_1:=1+\xi_1$ and iteratively define
    \[
    \delta_t:=1+\rho^2+\xi_t-\frac{\rho^2}{\delta_{t-1}}, \quad \hbox{for}\quad
    t=2, \ldots, T-1.
    \]
    Finally, define
    \[
    \delta_t:=1+\xi_t-\frac{\rho^2}{\delta_{t-1}}.
    \]
   \item[iv.] Define
   \[
   u_T=\frac{1}{\delta_T v_T},
   \] and iteratively define
    \[
    u_t:=\frac{\rho}{\delta_t}u_{t+1}, \quad \hbox{for}\quad
    t=T-1, \ldots, 1.
    \]
  \end{enumerate}
\end{definition}

Simple calculation shows that $d_t>0$ for $t=2,\ldots, T-1$ so that the sequences $\left(v_1, \ldots, v_T\right)$ and $\left(u_1, \ldots, u_T\right)$ in Definition \ref{def.2} are well-defined.
The following lemma provides some more results on the sequences $\left(v_1, \ldots, v_T\right)$ and $\left(u_1, \ldots, u_T\right)$.
\begin{lemma}\label{lem.1}
  For the given $\xi_1, \ldots, \xi_t>0$ and $\rho\in(-1,1)$, we have the following results on the sequences $\left(v_1, \ldots, v_T\right)$ and $\left(u_1, \ldots, u_T\right)$ in Definition \ref{def.2}.
  \begin{enumerate}
    \item[i.] $\left( v_{1:T}\right)$ is a decreasing sequence---$v_t\ge v_{t+1}$ for $t=1, \ldots, T-1$ where the equality holds for $\rho=0$.
    \item[ii.] $\left( u_{1:T}\right)$ is an increasing sequence---$u_t\le u_{t+1}$ for $t=1, \ldots, T-1$ where the equality holds for $\rho=0$.
  \end{enumerate}
\end{lemma}
\begin{proof}
To prove part i, it suffices to show that $\rho / d_t \leq 1$ for $t=2,\ldots, T$ by induction. First, one can easily see that $\rho / d_T = \rho / (1+\xi_T) \leq 1$. Further, if $\rho / d_{k+1} \leq 1$, then
$$
d_k-\rho = 1 +\rho^2-\rho+\xi_k-\frac{\rho^2}{d_{k+1}} = (1-\rho)^2 +\xi_k + \rho \left( 1-\frac{\rho}{d_{k+1}} \right) \geq 0,
$$
which lead to $d_t \geq \rho$ for $t=2, \ldots, T-1$ and subsequently $\rho / d_t \leq 1$ for $t=2, \ldots, T-1$.

For part ii, one can prove that $\rho / \delta_t \leq 1$ for $t=1, \ldots, T-1$ in the same manner.
\end{proof}

We also provide new symbols and related properties that are useful in the calculation of credibility factors.

\begin{definition}\label{def.3}
Under the distributional assumption in Model \ref{mod.1},
  define a $T\times T$ diagonal matrix $\boldsymbol{Q}_T^*$ by
  \[
  \boldsymbol{Q}_T^*:={\rm diag}\left( q_1^*, \cdots, q_T^*\right)
  \],
  where $q_t^*:=\psi\E{V\left( \lambda_t R_t\right)}$.
 We further define a $T\times T$ matrix $\boldsymbol{\Sigma}_T^*$ by
 \[
 \left[ \boldsymbol{\Sigma}_T^*\right]_{i,j}=
 \begin{cases}
   1+\xi_j, & (i,j)=(1,1)\quad\hbox{or}\quad(i,j)=(T,T);\\
   1+\rho^2+\xi_j, & i=j, \quad (i,j)\neq(1,1)\quad\hbox{and}\quad(i,j)\neq(T,T);\\
   -\rho, & j=i+1\quad\hbox{or}\quad j=i-1;\\
   0, &\hbox{otherwise.}
 \end{cases}
 \]
\end{definition}
Note that the matrix $\boldsymbol{\Sigma}_T^*$ satisfies
\begin{equation}\label{eq.a1}
\begin{aligned}
  \boldsymbol{\Sigma}_T^*&=\left( 1-\rho^2\right) \left( \boldsymbol{\Sigma}_{T, \rho} \right)^{-1}+
  \left(1-\rho^2\right)\sigma^2\boldsymbol{\Lambda}_T
  \left(\boldsymbol{Q}_T^*\right)^{-1}
  \boldsymbol{\Lambda}_T\\
  &=\left( 1-\rho^2\right) \left( \boldsymbol{\Sigma}_{T, \rho} \right)^{-1}+\boldsymbol{\Xi}_T.
\end{aligned}
\end{equation}

\begin{lemma}\label{lem.3}
  Under the settings in Model \ref{mod.1}, we have
  \begin{enumerate}
    \item[i.] For a $T\times T$ matrix $\boldsymbol{\Sigma}_{T, \rho}$
    \begin{equation}\label{eq.31}
    \left( 1-\rho^2 \right)\left( \boldsymbol{\Sigma}_{T, \rho}\right)^{-1}=
 \begin{cases}
   1, & (i,j)=(1,1)\quad\hbox{or}\quad(i,j)=(T,T);\\
   1+\rho^2, & i=j, \quad (i,j)\neq(1,1)\quad\hbox{and}\quad(i,j)\neq(T,T);\\
   -\rho, & j=i+1\quad\hbox{or}\quad j=i-1;\\
   0, &\hbox{otherwise.}
 \end{cases}
    \end{equation}
    and
    \begin{equation}\label{eq.32}
    \left( \boldsymbol{\Sigma}_{T, \rho}\right)^{-1} \left(\rho^T, \cdots, \rho^1 \right)'
    =\left(0, \cdots, 0, \rho \right)'
    \end{equation}
 \item[ii.] Moreover,
 \[
 \left(\boldsymbol{I}_T +\frac{1}{\sigma^2}\boldsymbol{Q}_T^*\left( \boldsymbol{\Lambda}_T\right)^{-1}
 \left( \boldsymbol{\Sigma}_{T, \rho}\right)^{-1}\left( \boldsymbol{\Lambda}_T\right)^{-1}
 \right)^{-1}=\sigma^2\left( 1-\rho^2\right)\boldsymbol{\Lambda}_T
 \left( \boldsymbol{\Sigma}_{T}^*\right)^{-1}\boldsymbol{\Lambda}_T
 \left( \boldsymbol{Q}_T^*\right)^{-1}.
 \]
  \end{enumerate}
\end{lemma}
\begin{proof}
  The proof of \eqref{eq.31} is from the well-known result on the inverse of AR(1) matrix $\boldsymbol{\Sigma}_{T, \rho}$. The result in \eqref{eq.32} is the immediate result of \eqref{eq.31}.
  The proof of part ii is from \eqref{eq.a1} and the simple matrix algebra.
\end{proof}

Finally, we present the analytical expression for the credibility premium under the distributional assumption in Model \ref{mod.1}.
\begin{theorem}\label{thm.1}
  Consider a state-space model comprising a state process $(R_t:t=0,1, \ldots)$ and observable time series $(Y_t:t=1,2, \ldots)$ in Model \ref{mod.1}. Subsequently, the credibility premium defined in \eqref{eq51} has the following property.
  \begin{enumerate}
    \item[i.] $\widehat{\alpha}_0^*,\cdots, \widehat{\alpha}_T^*$ in \eqref{eq.161} can be expressed as
        \[
        \left(\widehat{\alpha}_1^*,\cdots, \widehat{\alpha}_T^*  \right)=
        \rho\left(1-\rho^2\right)\sigma^2\lambda_{T+1} v_T \left(
                     u_1\frac{\lambda_1^2}{q_1^*},
                     \cdots ,
                     u_T \frac{\lambda_t^2}{q_T^*}
                 \right)
        \]
        and
        \[
        \widehat{\alpha}_0^*=1-\sum\limits_{t=1}^{T}\widehat{\alpha}_t^*.
        \],
        where $(u_{1:T})$, $v_T$ are defined in Definition \ref{def.2}, and $(q^*_{1:T})$ is defined in Definition \ref{def.3}.
    \item[ii.] The credibility premium defined in \eqref{eq51} is regular.

  \end{enumerate}
\end{theorem}
\begin{proof}
For convenience, we first show the representation for $\left(\widehat{\alpha}_1, \cdots, \widehat{\alpha}_T \right)$. Subsequently, we can use the equation in \eqref{eq.161}. Following the classical procedure in \citet{buhlmann2006course}, we have
\begin{equation}\label{eq.p0}
\begin{aligned}
\left(\widehat{\alpha}_1, \cdots, \widehat{\alpha}_T \right)'&=
\left( \boldsymbol{\Sigma}_T\right)^{-1} \left(\cov{Y_1, Y_{T+1}}, \cdots, \cov{Y_T, Y_{T+1}} \right)'\\
&=\sigma^2\lambda_{T+1}
\left( \boldsymbol{Q}_T^* + \sigma^2\boldsymbol{\Lambda}_t\boldsymbol{\Sigma}_{T, \rho}\boldsymbol{\Lambda}_T \right)^{-1}
\boldsymbol{\Lambda}_T\left( \rho^T, \cdots, \rho^1\right)'
\end{aligned}
\end{equation}
and
\[
\widehat{\alpha}_0=\lambda_{T+1} \left(1-\sum\limits_{t=1}^{T}\widehat{\alpha}_t \right).
\]
For the further derivation of \eqref{eq.p0}, we first investigate the auxiliary result on a symmetric tridiagonal matrix $\boldsymbol{\Sigma}_T^*$. Specifically, we have the following result on  the inverse of symmetric tri-diagonal matrix \citep{meurant1992review}
  \begin{equation}\label{eq.p1}
  \begin{aligned}
  \left(\boldsymbol{\Sigma}_t^*\right)^{-1}    \left(
      \begin{array}{c}
        0 \\
        \vdots \\
        0 \\
        1 \\
      \end{array}
    \right)
  &=\left(
      \begin{array}{ccccc}
        u_1v_1 & u_1v_2 & u_1v_3 & \cdots & u_1v_T \\
        u_1v_2 & u_2v_2 & u_2v_3 & \cdots & u_2v_T \\
        u_1v_3 & u_2v_3 & u_3v_3 & \cdots & u_3v_T \\
        \vdots & \vdots & \vdots & \ddots & \vdots \\
        u_1v_T & u_2v_T & u_3v_T & \cdots & u_Tv_T \\
      \end{array}
    \right)
    \left(
      \begin{array}{c}
        0 \\
        \vdots \\
        0 \\
        1 \\
      \end{array}
    \right)=v_T\left(
           \begin{array}{c}
             u_1 \\
             \vdots \\
             u_T \\
           \end{array}
         \right)
  \end{aligned}
  \end{equation}
  where $\left( u_{1:T}\right)$ and $\left( v_{1:T}\right)$ are defined in Definition \ref{def.2}.

Finally, we have
\[
\begin{aligned}
&\left(\widehat{\alpha}_1, \cdots, \widehat{\alpha}_T \right)'\\
&=\sigma^2\lambda_{T+1}
\left( \boldsymbol{Q}_T^* + \sigma^2\boldsymbol{\Lambda}_T\boldsymbol{\Sigma}_{T, \rho}\boldsymbol{\Lambda}_T \right)^{-1}
\boldsymbol{\Lambda}_T\left( \rho^T, \cdots, \rho^1\right)'\\
&=\sigma^2\lambda_{T+1}\left( \boldsymbol{Q}_T^*\right)^{-1}
\left(
\boldsymbol{I}_T+\sigma^2 \boldsymbol{\Lambda}_T\boldsymbol{\Sigma}_{T, \rho}\boldsymbol{\Lambda}_T
\left( \boldsymbol{Q}_T^*\right)^{-1}
\right)^{-1}
\boldsymbol{\Lambda}_T
\left( \rho^T, \cdots, \rho^1\right)'\\
&=\lambda_{T+1}\left( \boldsymbol{Q}_T^*\right)^{-1}
\left(
\boldsymbol{I}_T+\frac{1}{\sigma^2}\boldsymbol{Q}_T^*\left(\boldsymbol{\Lambda}_T\right)^{-1}
\left(\boldsymbol{\Sigma}_{T, \rho}\right)^{-1}\left(\boldsymbol{\Lambda}_T\right)^{-1}
\right)^{-1}
\boldsymbol{Q}_T^*
\left(\boldsymbol{\Lambda}_T\right)^{-1}\left(\boldsymbol{\Sigma}_{T, \rho}\right)^{-1}
\left( \rho^T, \cdots, \rho^1\right)'\\
&=\rho\sigma^2\left( 1-\rho^2 \right) \lambda_{T+1}\left( \boldsymbol{Q}_T^*\right)^{-1} \boldsymbol{\Lambda}_T \left(\boldsymbol{\Sigma}_{t}^*\right)^{-1}
\left( 0, \cdots, 0, 1\right)'\\
&=        \rho\left(1-\rho^2\right)\sigma^2\lambda_{T+1} v_T \left(
                     u_1\frac{\lambda_1}{q_1^*},
                     \cdots ,
                     u_T \frac{\lambda_T}{q_T^*}
                 \right),
\end{aligned}
\]
where the third equality is from Woodbury matrix identity \citep{hager1989updating, higham2002accuracy}, the fourth equality is from Lemma \ref{lem.3}, and the final equality is from \eqref{eq.p1}. Now,  \eqref{eq.161} concludes the proof of part i, and part ii immediately comes from part i.
\end{proof}

Finally, the proof of Theorem \ref{thm.2} follows immediately from Theorem \ref{thm.1}.

\section{A heterogeneous INAR(1) model} \label{sec.2.3.1}

Here, we investigate the ordering of credibility factors in
the heterogeneous  integer-valued autoregressive (INAR(1)) model proposed in \citet{gourieroux2004heterogeneous}. Owing to its simple and intuitive structure, INAR(1) model is widely used in the modelling of the frequency in insurance literature \citep{boucher2008models}.

For a non-negative integer-valued random variable $Y$ and the constant $p\in[0,1]$, define the thinning operator $p \circ Y$ as follows: conditional on $Y$, $p \circ Y$ is a random variable distributed by ${\rm BN}\left( Y, p\right)$---the binomial distribution with size $Y$ and probability $p$.

\begin{definition}[baseline or homogenous INAR(1) model]
Consider an autoregressive count model $\left(Y_{t}; t=1, 2, \cdots \right)$ such that $Y_1 \sim {\rm Pois}(\frac{\lambda}{1-p})$, and for all $t >1$,
\begin{equation}
\label{inar}
Y_t=p\circ Y_{t-1}+\epsilon_t
\end{equation}
where $\epsilon_t$'s are i.i.d. observations drawn from ${\rm Pois}(\lambda)$. Furthermore, conditionally on $Y_{t-1}$, the random variable $p\circ Y_{t-1}$ is assumed to be independent from the sequence $\epsilon_t$. It has been shown by \citep{mckenzie1985some, al1987first} that this process is stationary and has AR(1)-like ACF as, by definition, $E[Y_t|Y_{t-1}]=p Y_{t-1}+\lambda$.
\end{definition}

\begin{definition}[Heterogeneous INAR(1) model]\label{def.4}
The heterogeneous INAR(1) model extends the above baseline INAR(1) model by allowing the parameter $\lambda$ to be heterogeneous across the population. More precisely, we say that $\left( Y_t; t=1, 2, \cdots\right)$ is heterogeneous INAR(1), if
\begin{enumerate}
  \item[$i.)$]  given the unobservable heterogeneity factor $R$, the sequence $\left( Y_t; t=1, 2, \cdots\right)$ follows the INAR(1) process in \eqref{inar},
  where conditional on the past observation $Y_{t-1}, Y_{t-2}, \cdots, Y_1$, variables $p\circ Y_{t-1}$ and $\epsilon_t$ are independent with distributions ${\rm BN}\left( Y_{t-1}, p\right)$ and ${\rm Pois}\left( \lambda R \right)$, respectively.
  \item[$ii.)$] The marginal distribution of $R$ is ${\rm Gamma}\left( 1, \psi_0\right)$.
\end{enumerate}
\end{definition}

For the heterogeneous $INAR(1)$ in Model \ref{mod.1}, as $\E{R}=1$, we still have
\[
\E{Y_t}=\frac{\lambda}{1-p}.
\]
Moreover,
\begin{equation}\label{eq.12}
\cov{Y_t, Y_{t+h}}=\frac{\lambda}{1-p}\left( p^h +\frac{\lambda}{1-p}\psi_0 \right)
\quad\hbox{and}\quad
\Var{Y_t}=\frac{\lambda}{1-p}\left( 1 +\frac{\lambda}{1-p}\psi_0 \right)
\end{equation}
Consequently, it satisfies the positive covariance ordering in \eqref{eq.2}.
However, from Lemma \ref{lem.4} below, we have the ordering of credibility factors in \eqref{eq.17}
showing that the credibility factors do not satisfy the increasing property in \eqref{eq.16}.

%
%

\begin{lemma}\label{lem.4}
  Under the heterogeneous INAR(1) model in Definition \ref{def.4}, the credibility premium in \eqref{eq.10} 
is obtained as
\begin{equation}\label{eq.51}
\widehat{\alpha}_0=\frac{\lambda\left(1-p^2\right)}{(1-p)\left(b(t-p(t-2))+1+p\right)}\quad\hbox{and}\quad
\widehat{\alpha}_1=\frac{b(1-p)}{b(t-p(t-2))+1+p}
\end{equation}
and
\begin{equation}\label{eq.52}
\widehat{\alpha}_j=
\begin{cases}
(1-p)\widehat{\alpha}_1, & j=2, \cdots, t-1;\\
\widehat{\alpha}_1+p, &j=t;\\
\end{cases}
\end{equation}
where
\[
b=\frac{\lambda \psi_0}{1-p}.
\]

\end{lemma}
\begin{proof}
First, we calculate a covariance matrix, $\boldsymbol{\Sigma}_t$, of $\left( Y_1, \cdots, Y_t\right)$ as
\begin{equation}\label{eq.18}
\boldsymbol{\Sigma}_t=\left( \frac{\lambda}{1-p}\right)^2\psi_0 \boldsymbol{E}_t+ \frac{\lambda}{1-p}\boldsymbol{\Sigma}_{t, p}.
\end{equation}
Furthermore, by Sherman–Morrison formula \citep{sherman1950adjustment, press2007numerical}, observe the following matrix calculation
\begin{equation}\label{eq.19}
\left(\boldsymbol{\Sigma}_{t, p} + c \boldsymbol{E}_t \right)^{-1}=
\left(\boldsymbol{\Sigma}_{t, p}\right)^{-1} {-}\frac{c}{1+c \boldsymbol{1}_t^{\mathrm T}\left(\boldsymbol{\Sigma}_{t, p}\right)^{-1} \boldsymbol{1}_t}
\left(\boldsymbol{\Sigma}_{t, p}\right)^{-1}\boldsymbol{E}\left(\boldsymbol{\Sigma}_{t, p}\right)^{-1},
\end{equation}
where
\[
c=\frac{\lambda}{1-p}\psi_0.
\]
Subsequently, by \eqref{eq.15}, \eqref{eq.18}, and \eqref{eq.19},
the credibility premium in \eqref{eq.10} 
is obtained as in \eqref{eq.51} and \eqref{eq.52}.
\end{proof}

\section{Results on the ordering of the credibility factors under specific variance functions} \label{app.b}
\begin{corollary}\label{app.thm.1}
   Consider the time series $(Y_t)$ in the state-space model in Model \ref{mod.1}. Subsequently, we have the following results regardless of the choice of $(\lambda_t)$.
   \begin{enumerate}
     \item[i.] If we are given the following variance function
     \[
     V(x)=x
     \]
     of the EDF in \eqref{eq.ahn.1},
     then the credibility factors satisfy
        \[
        \widehat{\alpha}_1\le \cdots \le \widehat{\alpha}_T,
        \]
        where the equality holds if and only if $\rho=0$.
     \item[ii.] If we are given the following variance function
     \[
     V(x)=x^2
     \]
     of the EDF in \eqref{eq.ahn.1},
     then the credibility factors corresponding to the standardized observations satisfy
        \[
        \widehat{\alpha}_1^*\le \cdots \le \widehat{\alpha}_T^*,
        \]
        where the equality holds if and only if $\rho=0$.
   \end{enumerate}
\end{corollary}

The unit variance functions of Poisson and gamma distributions are given by
$V(\mu)=\mu $ and $V(\mu)=\mu^2 $, respectively. The following examples show that the numerical experiments on the ordering of the credibility factors confirm the results in Corollary \ref{app.thm.1}.

\begin{example}\label{ex.8}
  In this example, under the settings in Example \ref{exam.3}, we calculate the credibility factors of the non-standard observations, $(Y_1, \cdots, Y_T)$ with same parameters in Section \ref{sec.5.2}. We note that the unit variance function of the Poisson distribution is given by the identity function. Consequently, Corollary \ref{app.thm.1} implies that the credibility factors of non-standard observations $(Y_1,\cdots, Y_T)$ are increasing regardless of the choice of $(\lambda_t)$ as shown in Table \ref{tab.20}.

  \begin{table}[h!]
\caption{Credibility factors of the non-standardized observations in Example \ref{ex.8}
  (The credibility factors are written in the unit of $0.001$.)} \label{tab.20}
\centering
\begin{tabular}{|c|c|c|c c c c c|}
  \hline
    & $\rho$ & $(\lambda_1, \cdots, \lambda_5)$ & $\widehat{\alpha}_1$ & $\widehat{\alpha}_2$ & $\widehat{\alpha}_3$ & $\widehat{\alpha}_4$ & $\widehat{\alpha}_5$\\
    \hline
    Case 1.a & 0.3 & $(1, 1, 1, 1, 1)$ & 0.167  &   0.809  &   3.999  &  19.785  &  97.894  \\
    Case 1.b & 0.3 & $(0.001, 0.01, 0.1, 1, 10)$ & 0.131  &   0.438  &   1.467  &   5.114  & 24.871   \\
    Case 1.c & 0.3 & $(10, 1, 0.1, 0.01, 0.001)$ & 0.131  &   2.430  &  12.384  &   44.442  &   149.765   \\
    \hline
    Case 2.a & 0.6 & $(1, 1, 1, 1, 1)$ & 6.172  &  13.578  &  31.847   & 75.594  & 179.815    \\
    Case 2.b & 0.6 & $(0.001, 0.01, 0.1, 1, 10)$ &   4.586  &   7.646    &  12.785  &  22.016  & 48.859 \\
    Case 2.c & 0.6 & $(10, 1, 0.1, 0.01, 0.001)$ & 4.586  &  32.102   &  85.300   &  165.793  &   291.383 \\
  \hline
\end{tabular}
\end{table}
\end{example}

\begin{example}\label{ex.9}
  In this example, under the settings in Example \ref{exam.4}, we calculate the credibility factors of the non-standard observations $(Y_1, \cdots, Y_T)$ and the standard observations $(Y_1^*, \cdots, Y_T^*)$, respectively, with $T=5$, $\sigma^2=0.5$, and $\psi=0.5$.

  As the unit distribution function of the Gamma distribution is given by $V(\mu)=\mu^2$, Corollary \ref{app.thm.1} implies that the credibility factors of non-standard observations $(Y_1^*,\cdots, Y_T^*)$ are increasing regardless of the choice of $(\lambda_t)$. Note that the credibility factors of standard observations are not necessarily increasing as shown in Table \ref{tab.200}.

  \begin{table}[h!]
\caption{Credibility factors of the standardized observations and non-standardized observations in Example \ref{ex.9}
  (The credibility factors are written in the unit of $0.001$.)} \label{tab.200}
\centering
\begin{tabular}{|c|c|c|c c c c c|}
  \hline
    & $\rho$ & $(\lambda_1, \cdots, \lambda_5)$ & $\widehat{\alpha}_1$ & $\widehat{\alpha}_2$ & $\widehat{\alpha}_3$ & $\widehat{\alpha}_4$ & $\widehat{\alpha}_5$\\
    \hline
    Case 1.a & 0.3 & $(1, 1, 1, 1, 1)$ & 0.134   &   0.716  &   3.916 &  21.429  &  117.279  \\
    Case 1.b & 0.3 & $(0.001, 0.01, 0.1, 1, 10)$ & 0.134  &   0.072  &   0.039  &   0.021  & 0.012    \\
    \hline
    \hline
    & $\rho$ & $(\lambda_1, \cdots, \lambda_5)$ & $\widehat{\alpha}_1^*$ & $\widehat{\alpha}_2^*$ & $\widehat{\alpha}_3^*$ & $\widehat{\alpha}_4^*$ & $\widehat{\alpha}_5^*$\\
    \hline
    Case 2.a & 0.3 & $(1, 1, 1, 1, 1)$ & 0.134  &  0.716  &  3.916   & 21.429  & 117.279     \\
    Case 2.b & 0.3 & $(0.001, 0.01, 0.1, 1, 10)$ &   0.000  &   0.001    &  0.004  &  0.021  & 0.117   \\
  \hline
\end{tabular}
\end{table}
\end{example}

\end{document}